\documentclass{aa}  

\usepackage{graphicx}
\usepackage[T1]{fontenc}
\usepackage[usenames,dvipsnames]{xcolor}
\usepackage{natbib}
\usepackage[normalem]{ulem}
\usepackage{txfonts}
\usepackage[backref=page, hyperfootnotes=true, hidelinks, colorlinks, citecolor=blue, linkcolor=blue, linktocpage, bookmarks=true]{hyperref}
\usepackage{footnote}
\usepackage{float}
\usepackage{tabularx}
\usepackage[bottom, flushmargin]{footmisc}

\newcolumntype{Y}{>{\centering\arraybackslash}X}

\begin{document}

   \title{Two's a crowd? Characterising the effect of photometric contamination on the extraction of the global asteroseismic parameter $\nu_{\text{max}}$ in red-giant binaries}
   \titlerunning{Characterising the effect of photometric contamination in red-giant binaries}


\author{
    S.~Sekaran\inst{\ref{ivs}}
    \and
    C.~Johnston\inst{\ref{ivs}}
    \and
    A.~Tkachenko\inst{\ref{ivs}}
    \and
    P.~G.~Beck\inst{\ref{paul1}, \ref{paul2}}
    \and
    A.~Pr\v sa\inst{\ref{villanova}}
    \and
    K.~M.~Hambleton\inst{\ref{villanova}}
}

\institute{
    Instituut voor Sterrenkunde (IvS), KU Leuven, Celestijnenlaan 200D, B-3001 Leuven, Belgium \\
    \email{sanjay.sekaran@kuleuven.be}
    \label{ivs}
    \and
    Instituto de Astrof\'{\i}sica de Canarias, E-38200 La Laguna, Tenerife, Spain
    \label{paul1}
    \and
    Departamento de Astrof\'{\i}sica, Universidad de La Laguna, E-38206 La Laguna, Tenerife, Spain
    \label{paul2}
    \and
    Villanova University, Dept.~of Astrophysics and Planetary Science, 800 Lancaster Ave, Villanova PA 19085
    \label{villanova}
    }

   \date{Received August 16, 2018; accepted March 20, 2019}

 
  \abstract
   {Theoretical scaling relations for solar-like oscillators and red giants are widely used to estimate fundamental stellar parameters. The accuracy and precision of these relations have often been questioned in the literature, with studies often utilising binarity for model-independent validation. However, it has not been tested if the photometric effects of binarity introduce a systematic effect on the extraction of the seismic properties of the pulsating component(s).}
   {In this paper, we present an estimation of the impact of a contaminating photometric signal with a distinct background profile on the global asteroseismic parameter $\nu_{\text{max}}$ through the analysis of synthetic red-giant binary light curves.}
   {We generated the pulsational and granulation parameters for single red giants with different masses, radii and effective temperatures from theoretical scaling relations and use them to simulate single red-giant light curves with the characteristics of \textit{Kepler} long-cadence photometric data. These are subsequently blended together according to their light ratio to generate binary red-giant light curves of various configurations. We then performed a differential analysis to characterise the systematic effects of binarity on the extraction of $\nu_{\text{max}}$.}
   {We quantify our methodological uncertainties through the analysis of single red-giant light curves, both in the presence and absence of granulation. This is used as a reference for our subsequent differential binary analysis, where we find that the $\nu_{\text{max}}$ extraction for red-giant power spectra featuring overlapping power excesses is unreliable if unconstrained priors are used. Outside of this scenario, we obtain results that are nearly identical to single-star case.}
   {We conclude that i) the photometric effects of binarity on the extraction of $\nu_{\text{max}}$ are largely negligible as long as the power excesses of the individual components do not overlap, and that ii) there is minimal advantage to using more than two super-Lorentzian components to model the granulation signal of a binary red-giant.}


   \maketitle

\section{Introduction}
\label{sec: intro}

The revolution in red-giant asteroseismology began with the advent of the CoRoT space mission \citep{Auvergne2009}, which delivered high-precision and high duty-cycle photometry. These data enabled the first detections of non-radial pulsations in red-giants (e.g. \citealt{DeRidder2009, Kallinger2010a}) and gave rise to the birth of galactic archaeology (e.g. \citealt{Miglio2013}). The CoRoT data revolution had paved the way for the \textit{Kepler} space mission \citep{Borucki2010}, producing data of an unprecedented precision with a four-year nominal duty cycle. These new high-precision data have prompted numerous studies to test and improve asteroseismic scaling relations. 

First introduced by \cite{Kjeldsen1995}, these theoretical relations enable the determination of the masses and radii of stars exhibiting solar-like pulsations in a minimally model-dependent manner. \cite{Kjeldsen1995} had originally deduced these relations by calculating the global asteroseismic parameters known as the frequency of maximum oscillation power ($\nu_{\text{max}}$) and the large frequency separation ($\Delta\nu$) from the fundamental parameters of a sample of stars with detected solar-like oscillations, including the Sun (used as the reference star). Scaling relations take the form of a power-law with the effective temperature ($T_{\text{eff}}$), $\nu_{\text{max}}$ and $\Delta\nu$ as inputs. It is most commonly formulated in the following manner, first introduced by \cite{Kallinger2010a}:

\begin{equation}
\hfill \frac{M}{M_{\odot}} = \left(\frac{\nu_{\text{max}}}{\nu_{\text{max, }\odot}}\right)^{3}\left(\frac{\Delta\nu_{\odot}}{\Delta\nu}\right)^{4}\left(\frac{T_{\text{eff}}}{T_{\text{eff, }\odot}}\right)^{\frac{3}{2}};\hfill
\label{eq: Mscaling}
\end{equation}

\begin{equation}
\hfill \frac{R}{R_{\odot}} = \left(\frac{\nu_{\text{max}}}{\nu_{\text{max, }\odot}}\right)\left(\frac{\Delta\nu_{\odot}}{\Delta\nu}\right)^{2}\left(\frac{T_{\text{eff}}}{T_{\text{eff, }\odot}}\right)^{\frac{1}{2}}.\hfill
\label{eq: Rscaling}
\end{equation}

\noindent $M$ and $R$ are the stellar mass and radius, and the quantities with the $\odot$ subscript refer to the solar reference values. 

The large frequency separation ($\Delta\nu$) is the average difference in frequency between modes of the same spherical degree ($\ell$) and consecutive radial orders ($n$). This quantity was first defined theoretically by \cite{Tassoul1980} as a consequence of the asymptotic approximation. The asymptotic large frequency separation ($\Delta\nu_{\text{as}}$), which is directly proportional to the mean density ($\overline{\rho}$) of the star, is related to the sound speed ($c_{\text{s}}$) and radius ($R$) of the star according to the following equation:

\begin{equation}
\hfill \Delta\nu_{\text{as}} = \left(2\int_{0}^{R}\frac{dr}{c_{\text{s}}}\right)^{-1}.\hfill
\label{eq: deltanu}
\end{equation}

\noindent However, the observational large frequency separation ($\Delta\nu_{\text{obs}}$) is typically determined from the \textbf{mean} difference between frequencies of the modes (typically radial) of the same $\ell$ but different $n$ (i.e. $\Delta\nu_{\text{obs}}=<\nu_{n\text{, }\ell+1}-\nu_{n\text{, }\ell}>$). This necessarily implies that the observational and asymptotic $\Delta\nu$ are not equivalent, although the differences are small for red giants (see \citealt{Mosser2013} for a detailed discussion on the observed vs asymptotic $\Delta\nu$).

The frequency of maximum power ($\nu_{\text{max}}$) is an observational characterisation of the pulsational envelope of the star and was first used by \cite{Brown1991} to characterise the pulsations of Procyon, referring to it as an 'envelope peak.' It it most commonly used to refer to the frequency center of the Gaussian that is used to approximate the power excess displayed by all solar-like pulsators. Interestingly, there have been a few studies in the literature indicating that a Gaussian may not be the most optimal function to characterise the power excess for certain stars (e.g. Procyon, see \citealt{Arentoft2008} for more details). However, the Gaussian still remains the most ubiquitously-used function to characterise the oscillation power excess for solar-like oscillators. 

\citet{Brown1991}, and later \cite{Kjeldsen1995}, theorised that $\nu_{\text{max}}$ was related to the acoustic wave cutoff frequency in an isothermal atmosphere, and therefore scales with the surface gravity ($g$) and the effective temperature ($T_{\text{eff}}$) according to the following approximation:

\begin{equation}
\hfill \nu_{\text{max}} \propto \frac{g}{\sqrt{T_{\text{eff}}}}.\hfill
\label{eq: numaxgteff}
\end{equation}

\noindent While studies such as \cite{Belkacem2011, Belkacem2013} provide a full theoretical formulation for $\nu_{\text{max}}$, it has yet to be adopted ubiquitously in the current literature and is therefore mostly regarded as an observational quantity.

The accuracy of these scaling relations has often been tested in the literature, most commonly through ensemble studies (e.g. \citealt{Chaplin2011, Chaplin2014, Huber2011, Kallinger2010a, Kallinger2014}). These studies are complemented by the Tycho-\textsc{gaia} astrometric solution (TGAS) parallaxes \citep{Salgado2017}, an offshoot of the data products provided by the recently launched \textsc{gaia} space mission \citep{Gaia2016}. These parallaxes enable independent radii measurements that can be confronted to the scaling relation values, as demonstrated in \cite{DeRidder2016} and \cite{Huber2017}. 

Another method to test and verify scaling relations is through the confrontation of scaling relation-derived masses and radii with those extracted from eclipsing binary dynamics (e.g. \citealt{Frandsen2013, Gaulme2016,Brogaard2018,Themessl2018}), or otherwise constrained from spectroscopic binaries (e.g. \citealt{Beck2018a}). A notable example of the advantages offered by binarity is presented in \cite{Bellinger2017}, where the application of binary constraints allowed for the successful execution of inversion techniques in the modelling of 16~Cyg A and B. The common occurrence of binaries with solar-like pulsating components has allowed for the critical evaluation of evolutionary models \citep{Mathur2013,Appourchaux2015,Beck2014,White2017,Beck2018a,Li2018} and tidal theory \citep{Beck2018b}. The far less-common occurrence of eclipsing binaries with a solar-like pulsating red giant, however, requires more attention to detail to fully exploit the potential provided by dynamical mass and radius determinations.

The task of comparing asteroseismic and dynamical parameters is a complex one, with a number of requirements and drawbacks: For example, i) long time-series of both photometric, and high-resolution, high signal-to-noise spectroscopic observations (implying that said objects must be bright enough to be observed from the ground); ii) small sample sizes (few objects display eclipses of sufficient depth for analysis); and iii) complicated multi-step analyses involving binary and asteroseismic modelling (see e.g. \citealt{Rawls2016b}). However, precision and accuracy ($\sim$1\% uncertainty; \citealt{Torres2010}) of the masses and radii derived from eclipse modelling provide highly stringent constraints that scan be used to confront red-giant asteroseismic results, as exemplified by studies such as \cite{Gaulme2013,Gaulme2014,Gaulme2016}, \cite{Beck2014,Beck2018a}, \cite{Huber2015}, \cite{Brogaard2016}, \cite{Brogaard2018} and \cite{Themessl2018}. 

Of particular concern are the claims of \cite{Gaulme2016}, who had found an average overestimation of $\sim$15\%/5\% of red-giant mass/radius extracted from scaling relations when confronted with the mass/radius extracted from eclipsing binary dynamics. The departure of observationally-derived red-giant global asteroseismic parameters ($\nu_{\text{max}}$ and $\Delta\nu$) from theoretical ones is typically attributed to deficiencies in the scaling relations themselves. This phenomenon was most notably discussed in \cite{Mosser2013}, who had proposed empirical corrections to account for this discrepancy. Indeed, many attempts to propose corrections to scaling relations have been published (e.g. \citealt{White2011,Miglio2012,Guggenberger2016,Guggenberger2017,Sharma2016,Rodrigues2017}), typically involving new reference values for $\nu_{\text{max, }\odot}$ and $\Delta\nu_{\odot}$, or empirical or model-dependent corrections to the derived $\nu_{\text{max}}$ and $\Delta\nu$, or both. In fact, the aforementioned studies of \cite{Brogaard2018} and \cite{Themessl2018} also attempt to reconcile the discrepancy between eclipsing binary-analysis derived and scaling-relation derived parameters through similar corrections. However, there is still no agreement on the exact form of the corrections to the scaling relations. 

Most recently, non-linear scaling relations that claim to solve the eclipsing binary-analysis derived and scaling-relation derived parameter discrepancy have been proposed by \cite{Kallinger2018}, based on the systems studied by \cite{Gaulme2016}. However, this study, as well as the studies proposing scaling-relation corrections, still utilise the global asteroseismic parameter $\nu_{\text{max}}$ as an input parameter. So far, there have not been any studies that have attempted to quantify, in general for a wide range of stars, the impact (if any exists) that the light contribution of each component in a binary has on the extraction of these parameters. 

This question can then be further generalised: what is the impact of a contaminating photometric signal (whether a background object or companion) with a distinct background profile on the extraction of global asteroseismic parameters? This knowledge will be of particular importance in the context of the recently-launched \textsc{tess} space mission \citep{Ricker2015}, where the larger pixel sizes imply an increased likelihood of contamination from a background object. In this study, we attempt to answer the aforementioned question for the context of red-giant/red-giant binaries. This synergises with one of the suggestions in \cite{Beck2018a}: reformulating scaling relations in terms of mass ratios and light factors in order to exploit the typical outputs of binary analysis. To that end, we simulated the light curves of red-giant/red-giant binaries, with the characteristics of \textit{Kepler} long-cadence photometric data, and extracted the values of the global asteroseismic parameter $\nu_{\text{max}}$ from our synthetic light curves, comparing them to the input values calculated from theoretical scaling relations.

Section \ref{sec: method} describes the methodology used in our study. Section \ref{sec: SingleRGs} details the results of our single red-giant simulations, and Section \ref{sec: BinaryRGs} details the results of the binary red-giant simulations. We end by discussing our results and presenting our conclusions in Section \ref{sec: conclusions}.

\section{Methodology}
\label{sec: method}

\subsection{Red-giant signal components}
\label{subsec: RGsignals}

The signals that constitute a typical Red Giant (RG) power spectral density (PSD) that is generated from \textit{Kepler} long-cadence photometric data can be separated into three distinct categories: i) stellar pulsation, ii) stellar background, and iii) noise. The proper extraction of individual pulsational frequencies requires, and is strongly dependent on, the proper fitting and removal of the stellar background and noise.

The noise in the PSD is comprised of a frequency-independent (white) photon-shot noise component and a frequency-dependent (coloured) instrumental noise component, where the instrumental noise is thought to be a result of aperture optimisation during the pointing of \textit{Kepler}. This can be represented by the following equation:

\begin{equation}
\hfill P_{\text{noise}}(\nu)=W+\frac{2\pi\alpha^{2}/\beta}{1+(\nu+\beta)^{2}}.\hfill
\label{eq: noise}
\end{equation}

\noindent $P_{\text{noise}}(\nu)$ is the PSD of the noise signal, $W$ is the white noise, and $\frac{2\pi\alpha^{2}/\beta}{1+(\nu+\beta)^{2}}$ is the coloured instrumental noise represented by a Lorentzian profile, with $\alpha$ and $\beta$ representing the amplitude and turnover frequency of the instrumental noise. \cite{Kallinger2014} had shown that in practice, the coloured noise component is negligible, and therefore the noise can be assumed to be completely white.  

The background signal, however, is not well-understood. The general consensus is that the background consists of frequency-dependent (coloured) components that have been surmised to be the result of granulation occurring at different length- and time-scales. The first attempt to model this background signal was performed by \cite{Harvey1985}, who used a super-Lorentzian (or Harvey) profile that differed from a regular Lorentzian in that it had a larger exponent in the denominator. Since then, there have been several attempts to more accurately represent the background. Most notably, \cite{Kallinger2014} tested several background models using \textit{Kepler} data and concluded that, at the data quality level provided by \textit{Kepler} for a sufficiently bright target (i.e. brighter than $K_{\text{p}}=12$, where $K_{\text{p}}$ is the \textit{Kepler} magnitude), a background model consisting of two super-Lorentzian profiles of the form

\begin{equation}
\hfill P_{\text{bg}}(\nu)=\eta^{2}(\nu)\left[\sum_{i}^{2}\frac{\xi a_{\text{i}}^{2}/b_{\text{i}}}{1+(\nu/b_{i})^{4}}\right],\hfill
\label{eq: gran}
\end{equation}

\noindent best reproduced the data. $P_{\text{bg}}(\nu)$ is the PSD of the background signal, and the sum of the two super-Lorentzian profiles $\sum\limits_{i}^{2}\frac{\xi a_{\text{i}}^{2}/b_{\text{i}}}{1+(\nu/b_{i})^{4}}$ represent granulation occurring at different timescales, each with their respective amplitudes ($a_{\text{i}}$) and turnover frequencies ($b_{\text{i}}$), and the normalisation factor\footnote{See \cite{Kallinger2014} for a detailed explanation and mathematical background for the inclusion of a normalisation factor in the super-Lorentzian profiles.} $\xi=2\sqrt{2}/\pi$. $\eta^{2}(\nu)$ is the sinc-squared attenuation factor that accounts for the suppression of the amplitude of the oscillations and background occurring close to the Nyquist frequency due to the finite integration time of \textit{Kepler} \citep{Garcia2011}.

The pulsational signal present in RGs is understood to be a signature of stochastically-excited pulsations, with frequency-dependent driving and damping rates (see \citealt{Samadi2015} for a full description). This results in the formation of modes, which are typically described by spherical harmonics, around discrete frequencies dictated by the boundary conditions of stellar structure. The modes that result from this frequency-dependent driving and damping can be modelled as damped harmonic oscillators, which manifest as simple Lorentzian functions in the Fourier domain. Therefore, we can model the PSD of this pulsational signal with a sum of $m$ Lorentzian functions:

\begin{equation}
\hfill P_{\text{puls}}(\nu)=\sum_{n}^{m}\frac{A_{n}^{2}/(\pi\Gamma_{n})}{1+4(\nu-\nu_{0 \ (n)}/\Gamma_{n})^{2}}.\hfill
\end{equation}

\noindent $P_{\text{puls}}(\nu)$ is the PSD of the pulsational signal, and $A_{n}$, $\Gamma_{n}$, and $\nu_{0 \ (n)}$ are the amplitude, width and central frequency of the $n^{\text{th}}$ Lorentzian profile respectively. The pulsational signal appears as a power excess in the PSD of a RG, and as mentioned is approximated by a Gaussian centred at $\nu_{\text{max}}$, with an amplitude and width of $A_{\text{puls}}$ and $\sigma_{\text{puls}}$ respectively. The PSD of the Gaussian that represents the pulsational envelope ($P_{\text{env}}$) can be calculated using the following equation:

\begin{equation}
\hfill P_{\text{env}}(\nu) = A_{\text{puls}}\left[\text{exp}\left(\frac{-(\nu-\nu_{\text{max}})^{2}}{2\sigma_{\text{puls}}^{2}}\right)\right].\hfill
\label{eq: pulsenv}
\end{equation}

The standard practice in the literature is to fit the noise and background together with this Gaussian approximation of the pulsational power excess in order to constrain the morphology (or slope) of the background. The background fit is then removed from the PSD, and the individual peaks extracted by fitting Lorentzians to the residual signal.

\subsection{Light curve synthesis}
\label{subsec: LCsynth}

To synthesise the light curve of a RG, we need to simulate both the pulsational signal and the background signal. To simulate the pulsational signal, we first have to compile sets of template modes with known frequencies, amplitudes and mode lifetimes. 

To that end, we created a grid of stellar models of RGs using the stellar evolution code \textsc{mesa} (revision 10348; \citealt{Paxton2011, Paxton2018}). We adopted an exponential diffusive overshooting scheme \citep{Herwig2000}, and a predictive mixing scheme as described in \cite{Paxton2018}, with abundances based primarily on the work of \cite{Asplund2009}, with more-current values for certain elements taken from \cite{Nieva2012} and \cite{Przybilla2013}. The full contents of the inlist used to generate our evolutionary models is specified in Appendix \ref{sec: MESAInlist}. The models were made to evolve until the ignition of core-helium burning, resulting in a grid of RG models with masses of 1.0 $M_{\odot}$ to 2.0 $M_{\odot}$ in steps of 0.2 $M_{\odot}$, and a $\nu_{\text{max}}$ of approximately 20 $\mu$Hz to 200 $\mu$Hz in steps of 10 $\mu$Hz for each mass value. These $\nu_{\text{max}}$ values were calculated from the $M$ and $R$ outputs of \textsc{mesa} by using the following equations (a result of Equations \ref{eq: deltanu} and \ref{eq: numaxgteff}) as follows:

\begin{equation}
\hfill \nu_{\text{max}} = \left(\frac{M}{M_{\odot}}\right)\left(\frac{R_{\odot}}{R}\right)^{2}\left(\frac{T_{\text{eff, }\odot}}{T_{\text{eff}}}\right)^{\frac{1}{2}}\nu_{\text{max, }\odot};\hfill
\label{eq: numaxscaling}
\end{equation}

\begin{equation}
\hfill \Delta\nu = \left(\frac{M}{M_{\odot}}\right)^{\frac{1}{2}}\left(\frac{R_{\odot}}{R}\right)^{\frac{3}{2}}\Delta\nu_{\odot}.\hfill
\label{eq: delnuscaling}
\end{equation}

The $\nu_{\text{max}}$ range was chosen in such a way as to cover a large portion of the Red Giant Branch (RGB) and an upper limit of 200 $\mu$Hz for $\nu_{\text{max}}$ was set to ensure that the pulsational power excesses were well within the Nyquist frequency threshold of 284 $\mu$Hz for \textit{Kepler} long-cadence data. The ranges of the $M$, $R$, $T_{\text{eff}}$ and $\nu_{\text{max}}$ values of our grid are listed in Table \ref{tab: grid}. This grid of RG models was then used as inputs for the stellar pulsation code \textsc{gyre} (revision 5.1; \citealt{Townsend2013}), which we used to compute non-rotating pulsational models in the adiabatic framework. These pulsational models comprise the frequencies and normalised mode inertias of the $\ell=0$, 1, 2 and 3 modes for each \textsc{mesa} evolutionary model. The contents of the inlist used to generate our pulsational models are specified in Appendix \ref{sec: GYREInlist}.

\begin{table}[t]
\caption{The ranges of $M$, $R$, $T_{\text{eff}}$ and $\nu_{\text{max}}$ of the RG models used for the creation of our synthetic single- and binary-RG light curves.}
\begin{center}
\begin{tabular}{ ccccc } 
 \hline
 \hline
  & $M$ [$M_{\odot}$] & $R$ [$R_{\odot}$] & $T_{\text{eff}}$ [K] & $\nu_{\text{max}}$ [$\mu$Hz]\\ 
 \hline
 Min & 1.0 & 4.2 & 4192 & 20.2\\ 
Max & 2.0 & 19.0 & 4913 & 202.6\\ 
 \hline
\end{tabular}
\end{center}
\label{tab: grid}
\end{table}

The corresponding mode amplitudes for each frequency were then generated using the following steps, similar to the methodology detailed in Section 4.3 of \cite{Kallinger2014} for the creation of synthetic RG data:

\begin{enumerate}
\item{Initial amplitude ratios ($A_{\text{ratio}}$) between 0 and 1 were assigned to each frequency ($\nu_{\text{i}}$) by using a Gaussian centred on $\nu_{\text{max}}$ with a standard deviation\footnote{We use a standard deviation of $1.5\Delta\nu$, resulting in a full-width at half-maximum (FWHM) of $3.5\Delta\nu$, which is well within the FWHM ranges for RGs reported by \cite{Mosser2010}.} of $1.5\Delta\nu$. This is represented by the following equation:

\begin{equation}
\hfill A_{\text{ratio}}=\text{exp}\left(\frac{-(\nu_{\text{i}}-\nu_{\text{max}})^{2}}{2(1.5\Delta\nu)^{2}}\right).\hfill
\end{equation}

} 
\item{These amplitude ratios were then modulated by mode visibilities of 1.00, 1.35, 0.64 and 0.071 for the $\ell=0$, 1, 2 and 3 modes respectively, as reported by \cite{Mosser2012b} for RGB stars.}
\item{The amplitude ratios of the $\ell=1$, 2 and 3 modes were then further modulated by the inverse of the local relative mode inertia\footnote{The normalised mode inertias exhibit local minima in intervals of $\Delta\nu$, but display a global decreasing trend that can span more than one order of magnitude (e.g. Figure \ref{fig: MI}). This results in modes with high amplitudes clustered around $\Delta\nu$, which mirrors observational oscillation patterns. See \cite{Dupret2009} for a full description.}.}
\item{To obtain realistic pulsational amplitudes, we refer to the characterisation of pulsational amplitudes reported in \cite{Corsaro2013}, where different amplitude scaling-relation models were investigated in a Bayesian framework. We calculated the total bolometric pulsational amplitude $A_{\text{puls, bol}}$ from the scaling relation given in Equation 19 of \cite{Corsaro2013}:

\begin{equation}
\begin{split}
\hfill A_{\text{puls, bol}}=d\left(\frac{\nu_{\text{max}}}{\nu_{\text{max, }\odot}}\right)^{2p-3q-r+0.2}&\left(\frac{\Delta\nu}{\Delta\nu_{\odot}}\right)^{4q-4p}\\
\cdot&\left(\frac{T_{\text{eff}}}{T_{\text{eff, }\odot}}\right)^{5p-1.5q-r+0.2}.\hfill
\end{split}
\end{equation}

\noindent The values for the coefficients and exponents in the equation are as follows: $d=e^{0.45}$, $p=0.602$, $q=1.31$, and $r=5.87$. These are taken from the long-cadence expectation values of model $\mathcal{M}_{4\text{, }d}$ of \cite{Corsaro2013}, which they found to have the best Bayesian Information Criterion (BIC) value in their analysis.}
\item{Amplitudes (in ppm) for each frequency can then obtained by taking the product of the modulated amplitude ratios and $A_{\text{puls, bol}}$.}
\end{enumerate}

We generated mode lifetimes for each frequency based on the work of \cite{Vrard2018}, who discussed both global (variation with stellar mass and temperature across all stars) and local (variation with frequency within a single star) modewidth variation in their study. Based on their results, we generated the modewidth at $\nu_{\text{max}}$ for each star by using the following equation:

\begin{equation}
\hfill \Gamma(\nu_{\text{max}}) = n M^{0.27}T_{\text{eff}}^{4.74}.\hfill
\label{eq: globalgamma}
\end{equation}

\noindent $\Gamma(\nu_{\text{max}})$ is the modewidth at $\nu_{\text{max}}$, with the exponents of $M$ and $T_{\text{eff}}$ taken from the characterisation of RGB stars by \cite{Vrard2018}. n is a multiplicative constant used to ensure that the modewidths occupy a range of 0.05 to 0.17 $\mu$Hz, which is well within the distribution of modewidths displayed in Figure 3 of \cite{Vrard2018}.

\cite{Vrard2018} also investigated local modewidth variation for the radial modes of RGs as per the work of \cite{Appourchaux2014} for subgiants and \cite{Lund2017} for dwarfs. While they found a general increase in modewidth with frequency, they did not find the modewidth 'dip' around $\nu_{\text{max}}$ (although they did find evidence of a reduction in the rate of increase around $\nu_{\text{max}}$). Although their results suffer from large uncertainties, we note a general exponential morphology of the variation of modewidths with frequency. As such, we chose to adopt an exponential scaling of local modewidth with frequency as follows:

\begin{equation}
\hfill \Gamma(\nu) = \Gamma(\nu_{\text{max}}) \left[\text{exp}\left(\frac{z(\nu-\nu_{\text{max}})}{\nu_{\text{max}}}\right)\right].\hfill
\label{eq: localgamma}
\end{equation}

\noindent $\Gamma(\nu)$ is the modewidth at frequency $\nu$, and $z=3.0$ is a constant designed to keep the modewidth increase with frequency within the ranges reported by \cite{Vrard2018}. 

\begin{figure}
\includegraphics[width=\hsize]{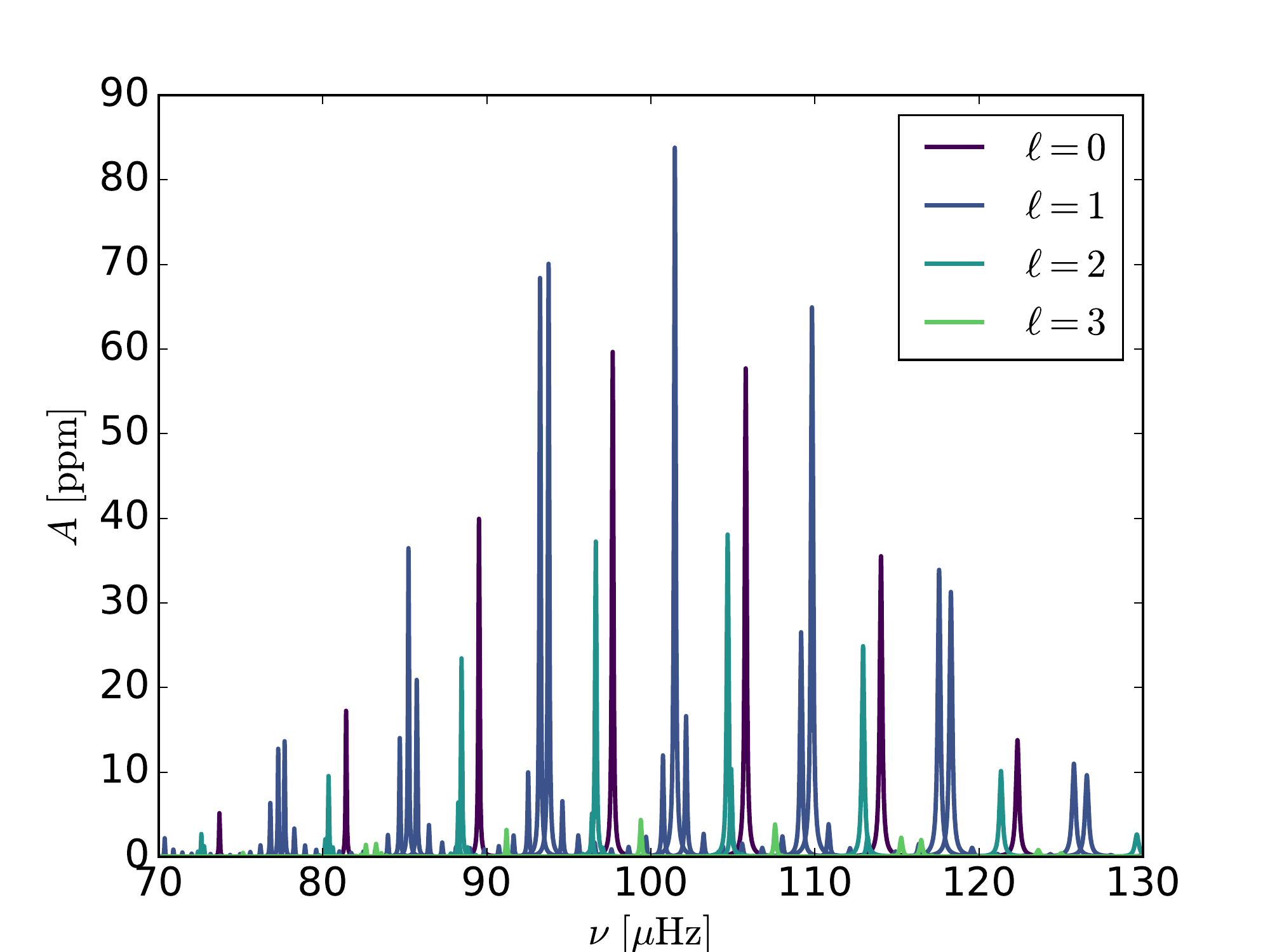}
\caption[An example of a synthetic RG pulsation spectrum.]{An example of a synthetic RG pulsation spectrum for a star with $\nu_{\text{max}}\sim100$ $\mu$Hz, used as an input to generates synthetic pulsational time-series for the RG light curves. The $\ell=0$, 1, 2, and 3 modes are represented by Lorentzians and colour-coded as displayed in the legend. The presence of multiplets, particularly around the $\ell=1$ modes, correspond to mixed modes with the same $p$-mode radial order but different $g$-mode radial order.}
\label{fig: PulsInp}
\end{figure}

The frequencies, amplitudes and modewidths (see Figure \ref{fig: PulsInp} for an example of a pulsation spectrum) are then fed to a first-order auto-regressive function to generate a synthetic time-series, following the methodology laid out in \cite{DeRidder2006}. This procedure transforms the 'perfect' comb of our input pulsation spectrum into a more-realistic representation of a RG pulsation spectrum (c.f. Figure \ref{fig: singlepuls_fitex}). The time sampling of this function was setup in such a way as to produce a time-series with the same time-stamps as a full four-year \textit{Kepler} long cadence light curve with no gaps.

To simulate the granulation signal for our light curves, we first generated super-Lorentzian profiles as per Equation (\ref{eq: gran}). The granulation amplitudes ($a_{\text{i}}$) and turnover frequencies ($b_{\text{i}}$) were calculated from the scaling relations and coefficients listed in Table 2 of \cite{Kallinger2014}, using the $M$ values from \textsc{mesa} and our calculated $\nu_{\text{max}}$ values:

\begin{equation}
\hfill a_{\text{i}} = k\nu_{\text{max}}^{s}M^{t};\hfill
\label{eq: granamp}
\end{equation}

\begin{equation}
\hfill b_{\text{i}} = k\nu_{\text{max}}^{s}.\hfill
\label{eq: grantime}
\end{equation}

\noindent The coefficients $k$, $s$, and $t$ for $a_{1\text{, }2}$ and $b_{1\text{, }2}$ are listed in Table \ref{tab: grancoeffs}. As the granulation signals of observational PSDs are often very noisy, we added a fixed amount of Gaussian noise to each granulation profile to make them more realistic.

\begin{table}[t]
\caption{The coefficients $k$, $s$, and $t$ for the granulation amplitudes ($a_{1\text{, }2}$) and turnover frequencies ($b_{1\text{, }2}$).}
\begin{center}
\begin{tabular}{ ccccc } 
 \hline
 \hline
  & $a_{1}$ & $a_{2}$ & $b_{1}$ & $b_{2}$\\ 
 \hline
 $k$ & 3710 & 2226 & 0.317 & 0.948\\ 
 $s$ & -0.613 & -0.613 & 0.970 & 0.992\\ 
 $t$ & -0.26 & -0.26 & -- & --\\ 
 \hline
\end{tabular}
\tablefoot{These coefficients were taken from Table 2 of \cite{Kallinger2014}.}
\end{center}
\label{tab: grancoeffs}
\end{table}

To mimic the procedure that we had performed for the pulsation spectrum, we chose to generate synthetic time-series for our granulation spectra. While it is possible to transform a simple Harvey profile into a time-series using an auto-regressive function, there is currently no analogue that can transform our granulation profiles into a time-series. We therefore performed inverse Fourier Transforms (iFTs) of our granulation PSDs in order to generate them. Similar to the pulsational time-series, the iFTs are calculated in such a way that the resulting time-series has the same time-stamps as a full \textit{Kepler} long-cadence light curve. 

We chose to include a luminosity-dependent white noise component as per the work of \cite{Pande2018}, who characterised the variation of white noise with magnitude for approximately 2100 stars from short-cadence \textit{Kepler} data. They reported an almost-linear variation of the logarithm of white noise with magnitude. We adapted this result and chose to scale the logarithm of white noise with the logarithm of luminosity as per the following equation:

\begin{equation}
\hfill \text{log }W = -\text{log }L+c .\hfill
\label{eq: whitescaling}
\end{equation}

\noindent $L$ is the luminosity of star, which is one of the outputs of \textsc{mesa}, and $c$ is a constant designed to ensure that the range of white noise amplitudes are within that of the stars in the 8 to 12 \textit{Kepler} magnitude range as reported by \cite{Pande2018}. This is within the range in which a two super-Lorentzian-component model can accurately describe the granulation signal of a \textit{Kepler} RG, as detailed in \cite{Kallinger2014}. The implicit assumption being made here is that these stars are all at the same distance, which we adopt in order to investigate the effect of white noise on the extraction of $\nu_{\text{max}}$. Similar to the granulation signal, iFTs were performed in order to produce white noise time-series.

The three time-series are then added to produce the final synthetic light curve for a single RG of mass $M$, radius $R$ and effective temperature $T_{\text{eff}}$.

\subsection{Synthetic PSD fitting}
\label{subsec: LCfit}

We fit the synthetic PSDs, generated by taking the Fourier transforms of our synthetic light curves, as per what is done in the literature (see Section \ref{subsec: RGsignals}). This fitting function, essentially a sum of the pulsational, granulation, and noise model functions, takes the following form:

\begin{equation}
\begin{split}
P_\text{model}(\nu)=W&+A_{\text{puls}}\left[\text{exp}\left(\frac{-(\nu-\nu_{\text{max}})^{2}}{2\sigma_{\text{puls}}^{2}}\right)\right]\\
&+\sum_{i}^{2}\frac{(2\sqrt{2}/\pi)\tau_{\text{gran, i}}A_{\text{gran, i}}^{2}}{1+(\nu\tau_{\text{gran, i}})^{4}}.
\end{split}
\label{eq: fit}
\end{equation}

\noindent The terms in the function echo those in Equations (\ref{eq: noise}), (\ref{eq: gran}) and (\ref{eq: pulsenv}): $P_{\text{model}}(\nu)$ is the PSD of the model; $W$ is the white noise; $A_{\text{puls}}\left[\text{exp}\left(\frac{-(\nu-\nu_{\text{max}})^{2}}{2\sigma_{\text{puls}}^{2}}\right)\right]$ represents the Gaussian power excess, with the amplitude, width and central frequency represented by $A_{\text{puls}}$, $\sigma_{\text{puls}}$ and $\nu_{\text{max}}$ respectively; and $\sum\limits_{i}^{2}\frac{(2\sqrt{2}/\pi)\tau_{\text{gran, i}}A_{\text{gran, i}}^{2}}{1+(\nu\tau_{\text{gran, i}})^{4}}$ are the two super-Lorentzian components representing the granulation signal, with the amplitude and characteristic timescale of the $i^{\text{th}}$ super-Lorentzian represented by $A_{\text{gran, i}}$ and $\tau_{\text{gran, i}}$ respectively.

We perform our fit in a Bayesian framework using the Markov Chain Monte Carlo (MCMC) routine \texttt{emcee} \citep{ForemanMackey2013}. \texttt{emcee} makes use of an ensemble sampler with numerous Markov chains to efficiently sample the posterior probability distribution as described by Bayes' theorem:

\begin{equation}
\hfill P(H_{\text{i}}|D,I)=\frac{P(H_{\text{i}}|I)P(D|H_{\text{i}},I)}{P(D|I)}.\hfill
\end{equation}

\noindent $P(H_{\text{i}}|I)$ is known as the prior probability, which is the probability of the hypothesis $H_{\text{i}}$ with knowledge of prior information $I$, but in the absence of the data $D$. $P(H_{\text{i}}|D,I)$ is known as the posterior probability, which is the probability of the hypothesis with knowledge of both $I$ and $D$. The quantities $P(D|H_{\text{i}},I)$ and $P(D|I)$ are the likelihood of $H_{\text{i}}$ and the global likelihood of all hypotheses respectively.

For numerical simplicity, we evaluate Bayes' theorem in log-space, where our log-likelihood function ($\mathcal{L}$) is

\begin{equation}
\hfill\text{ln }\mathcal{L}=-\sum_{v}\left(\text{ln }P_\text{model}(\nu)+\frac{P_\text{input}(\nu)}{P_\text{model}(\nu)}\right).\hfill
\end{equation}

\noindent $P_\text{input}(\nu)$ is the PSD of the input light curve that has been smoothed by a Gaussian filter. We use a smoothed PSD to mitigate the impact of noise in the background profile on the fit. We adopt a convergence criterion based on the integrated autocorrelation time ($\tau_{\text{autocorr}}$) as per \cite{Goodman2010}. We compute $\tau_{\text{autocorr}}$ for every 100 iterations and consider the MCMC routine to have achieved convergence if $\tau_{\text{autocorr}}$ changes by less than 1\% between two successive computations. 

\begin{figure}
\includegraphics[width=\hsize]{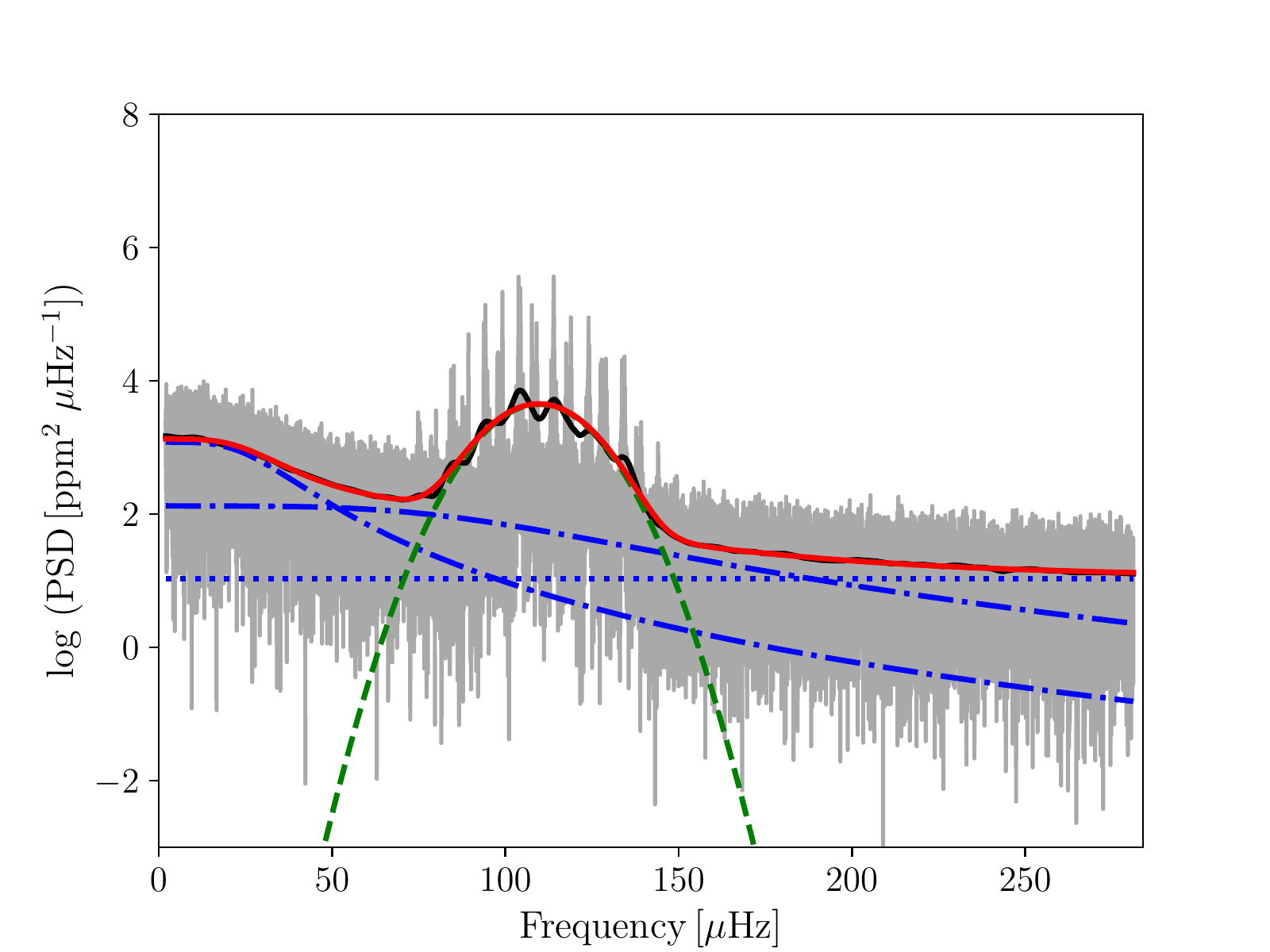}
\caption[An example of a single-RG PSD fit.]{An example of a single-RG PSD fit. The grey line is the unsmoothed PSD, the black line is the smoothed PSD, the red line represents the overall fit, the dashed green line represents the Gaussian used to fit the pulsational power excess, the blue dotted line represents the white noise, and the blue dashed lines represent the super-Lorentzians used to fit the granulation signal.}
\label{fig: singlegran_fitex_sec2}
\end{figure}

After our MCMC routines have achieved convergence, we extract our fit parameters and determine $\nu_{\text{max, fit}}$ from the marginalised posterior distributions, comparing them with our input $\nu_{\text{max, scaling}}$ from scaling relations. An example fit of the PSD of one of our synthetic light curves is shown in Figure \ref{fig: singlegran_fitex_sec2}.

\subsection{Red-giant binary simulation setup}
\label{subsec: BinLCsynthfit}

To create the light curve of a RG/RG binary, it was necessary to deal with the pulsations and granulation, and white noise separately. We first generated the white noise signal of the binary light curve according to Equation (\ref{eq: noise}) by using the total luminosity of the individual components as the input. We then scaled the pulsational and granulation signals of the individual components by their individual light contribution and added them together with the white noise according to the following equation:

\begin{equation}
\begin{split}
F_{\text{binary}}=\frac{1}{L_{\text{A}}+L_{\text{B}}}\bigg[&L_{\text{A}}\left(F_{\text{puls, A}}+F_{\text{gran, A}}\right)\\
+&L_{\text{B}}\left(F_{\text{puls, B}}+F_{\text{gran, B}}\right)\bigg]+W.
\end{split}
\end{equation}

\noindent $F_{\text{binary}}$ is the flux of the binary light curve, and $L_{\text{A}}$, $F_{\text{puls, A}}$, $F_{\text{gran, A}}$, $L_{\text{B}}$, $F_{\text{puls, B}}$, and $F_{\text{gran, B}}$ are the luminosities, and the pulsational and granulation fluxes of each binary component ($A$ and $B$) respectively. Implicit in this methodology is that all orbital and tidal effects are ignored\footnote{The observational signature and efficiency of dynamical and equilibrium tides in RG-binaries are discussed in \cite{Beck2018b}.}.

We created two sets of light curves with each set containing a specific pulsational configuration: i) one pulsating component (e.g. \citealt{Gaulme2014,Beck2014}), and ii) two pulsating components (e.g. \citealt{Rawls2016a,Beck2018a}). For each set of binary light curves, we tested the impact of different granulation models by performing fits with i) two super-Lorentzians (as in the single-RG case), and with ii) four super-Lorentzians (two super-Lorentzians for each component), to model the granulation signal of each binary light curve.

\section{Single red-giant simulations and results}
\label{sec: SingleRGs}

\begin{figure*}[t]
\centering
\includegraphics[width=\hsize]{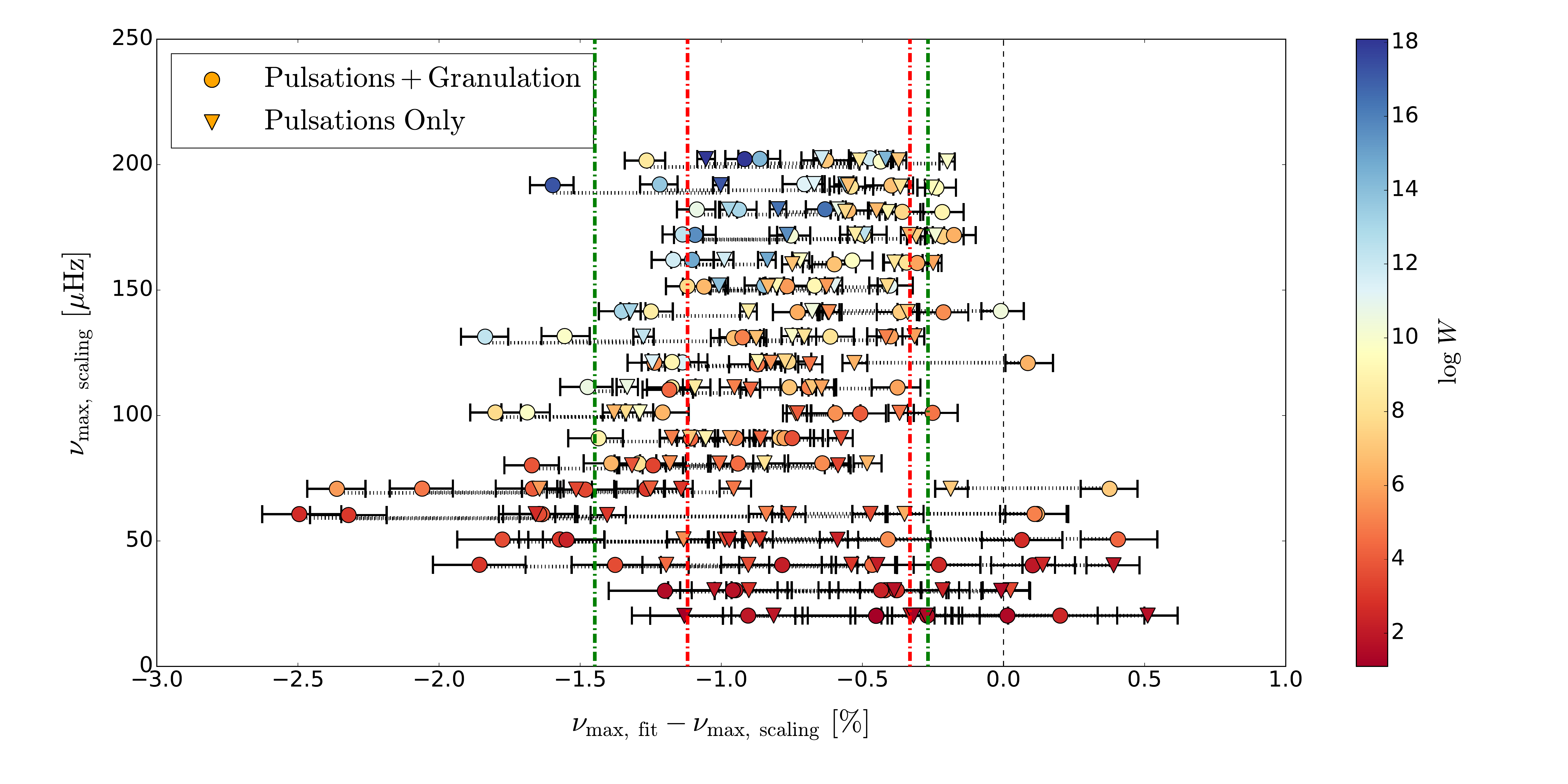}
\caption[A comparison of the $\nu_{\text{max}}$ values extracted from our synthetic single-RG PSDs.]{A comparison of the $\nu_{\text{max}}$ values extracted from our synthetic single-RG PSDs. We tested synthetic light curves containing only a pulsational signal (triangles), and both a pulsational and a granulation signal (circles). The vertical axis corresponds to the scaling relation ($\nu_{\text{max, scaling}}$) input values of the light curves, and the horizontal axis corresponds to the percentage difference between the fit ($\nu_{\text{max, fit}}$) and the scaling relation ($\nu_{\text{max, scaling}}$) values. The vertical dashed line represents the zero-point of the difference between $\nu_{\text{max, fit}}$ and $\nu_{\text{max, scaling}}$. The dash-dotted lines represent the 1-$\sigma$ level about the mean scatter of the datapoints corresponding to the pulsational signal (red), and both the pulsational and granulation signals (green). The horizontal dotted lines connect the datapoints corresponding to the same RG model. The error bars correspond to the 68\% Bayesian credible intervals of the marginalised posterior distributions of the fit parameters. The symbols are colour-coded according the logarithm of the white noise ($\text{log }W$) included in the light curves and PSDs.}
\label{fig: GranvPuls}
\end{figure*}

To ensure the self-consistency of our methodology and to establish a baseline for comparison with our subsequent binary results, we first simulated the light curves and fit the PSDs of the 114 single RGs that we created using our methodology. We investigated the effect of the granulation on the extraction of $\nu_{\text{max}}$ by performing one iteration of fits with only the pulsational signal (e.g. Figure \ref{fig: singlepuls_fitex}), and another with both the pulsational and granulation signals (e.g. Figure \ref{fig: singlegran_fitex}) included in the light curves of the RGs. This is demonstrated in Figure \ref{fig: GranvPuls}, showcasing the percentage difference between the fit ($\nu_{\text{max, fit}}$) and the scaling relation ($\nu_{\text{max, scaling}}$) input values. The mean percentage difference and the 1-$\sigma$ scatter of the $\nu_{\text{max, fit}}-\nu_{\text{max, scaling}}$ values, and the mean precision of the extraction of $\nu_{\text{max, fit}}$ are detailed in Table \ref{tab: SingleRG}. 

It can be seen that in general, for the iterations both with and without the inclusion of granulation in the PSDs, there is a systematic underestimation of the $\nu_{\text{max, fit}}$ compared to the $\nu_{\text{max, scaling}}$ values. This was the opposite of the expected result, which was a systematic \textbf{overestimation} due to an excess of power in the frequency domain above $\nu_{\text{max, scaling}}$ as the modewidths increase with increasing frequency. 

However, there is an additional source of asymmetry in the input pulsational spectra: uneven distribution of modes and mode amplitudes. For example, in the input pulsation spectrum displayed in Figure \ref{fig: PulsInp} (which correponds to a $\nu_{\text{max, scaling}}\sim100$ $\mu$Hz), there is a larger number of strong peaks in the frequency domain below $\nu_{\text{max, scaling}}$ than above it, particularly from the $\ell=1$ mixed-mode multiplets. This distribution is a result of the mode inertia distribution (e.g Figure \ref{fig: MI}) for the frequencies output by \textsc{gyre}, which directly affects the amplitudes of the modes that were used in the input. This result indicates that the input pulsational spectra generally have more power in the frequency domain below $\nu_{\text{max, scaling}}$ than above it, and is the source of our systematic uncertainty. To confirm that this is indeed the reason for the obtained systematic effect, we performed a test by artificially symmetrising our pulsational inputs with respect to $\nu_{\text{max, scaling}}$ and repeating the same fitting procedure. We found that the systematic offset between the $\nu_{\text{max, fit}}$ and $\nu_{\text{max, scaling}}$ was no longer present, confirming our suspicion that its origins was indeed in the uneven distribution of modes and amplitudes.

Another result from these simulations is the difference in the $\nu_{\text{max, fit}}$ values extracted from PSDs with and without granulation. The $\nu_{\text{max, fit}}$ values for the iteration with granulation are slightly more underestimated on average compared to the $\nu_{\text{max, scaling}}$ values, seemingly exacerbating the asymmetry in the power excesses. In addition, the scatter of the $\nu_{\text{max, fit}}$ is greater, and the extraction of the $\nu_{\text{max, fit}}$ values is less precise for the iteration with granulation. These phenomena can be explained by the stochastic nature of the granulation signal and the degenerate nature of the background fit, where a smooth function (the two super-Lorentzians) are being fit to a rather noisy background. The effect of the inclusion and the fitting of the granulation seems to be rather unpredictable as the degree of underestimation of $\nu_{\text{max, fit}}$ is decreased in some cases when comparing the results with and without granulation.

It can also be seen in Figure \ref{fig: GranvPuls} that the precision of the extraction seems to decrease as the $\nu_{\text{max, scaling}}$ values decrease, and that this decrease seems to be correlated with a decrease in the amount of white noise. However, this is simply due to the fact that we represented our results in terms of percentages: the absolute errors for the different stars within a single iteration (i.e. whether with or without granulation) were very similar and similarly uncorrelated with the amount of white noise in the PSDs. 

In addition, there seems to be a marked increase in the percentage difference and scatter of the $\nu_{\text{max, fit}}$ values, for the iterations with and without granulation, as the $\nu_{\text{max, scaling}}$ values decrease below $\sim$100 $\mu$Hz. While asymmetrical mode distributions and our representation of the errors in percentages are partially to blame, there seems to be a trend in the scatter below $\sim$100 $\mu$Hz that requires additional explanation. We posit that this result is a consequence of the large changes in the gradient of the granulation morphology (i.e. the 'flattening' of the background slope) below $\sim$50 $\mu$Hz (see e.g. Figure \ref{fig: singlegran_fitex}). Due to the fact that the pulsational power excesses lie partially on top of this 'flattened' portion of the granulation morphology, the extraction of $\nu_{\text{max, fit}}$ between the two iterations shows greater deviance.

Overall, we observe systematic errors of the order of 1\% in $\nu_{\text{max}}$, 3\% in $M$ and 1\% in $R$, with an intrinsic scatter of the order of 0.5\% in $\nu_{\text{max}}$, 1.5\% in $M$ and 0.5\% in $R$. For all three parameters, the combined errors are smaller or of the order of the typical uncertainties of 4\%/2\% in mass/radius reported for the scaling relations in literature (e.g. \citealt{PerezHernandez2016,Kallinger2018}).

The systematic errors and intrinsic scatter of the masses and radii follow from the scaling relations (Equations \ref{eq: Mscaling} and \ref{eq: Rscaling}) as expected, with $\Delta M = 3\cdot|\text{d(log }\nu_{\text{max}})|$ and $\Delta R = |\text{d(log }\nu_{\text{max}})|$. As such, in this and in the subsequent sections, we only present the mean percentage difference, 1-$\sigma$ scatter, and the mean precision of the $\nu_{\text{max}}$ values extracted from our PSDs, as those of the masses and radii can effectively be calculated from the $\nu_{\text{max}}$ results.

\begin{table}[!h]
\caption{The mean percentage difference, 1-$\sigma$ scatter and precision of the $\nu_{\text{max}}$ values extracted from our single-RG PSDs.}
\begin{center}
\begin{tabular}{ ccc }
\hline
\hline
\multicolumn{3}{c}{$\nu_{\text{max, fit}}-\nu_{\text{max, scaling}}$ [\%]}\\
 \hline
& \multicolumn{1}{c}{Pulsations Only} & \multicolumn{1}{c}{Pulsations+Granulation}\\
 \hline
 Mean & -0.7 & -0.9\\ 
1-$\sigma$ & 0.4 & 0.6\\
Precision & 0.1 & 0.2\\ 
 \hline
\end{tabular}
\tablefoot{The precision corresponds to the errors propagated from the 68\% Bayesian credible intervals of the marginalised posterior distributions of the fit parameters.}
\end{center}
\label{tab: SingleRG}
\end{table}

\section{Red-giant binary simulations and results}
\label{sec: BinaryRGs}

\begin{figure}[!t]
\includegraphics[width=\hsize]{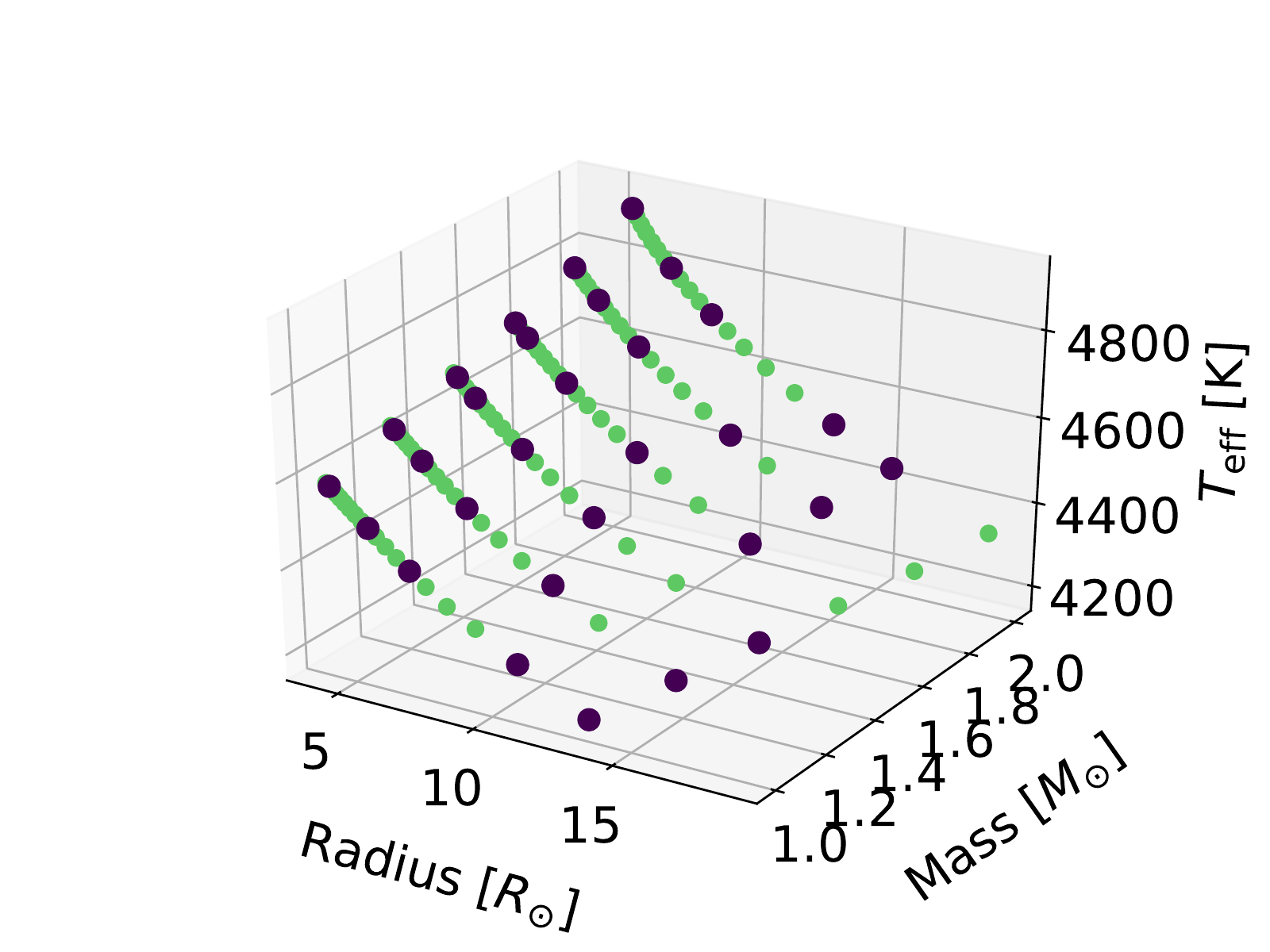}
\caption[The $M$, $R$, and $T_{\text{eff}}$ values of the individual components used to create our binary grid.]{The $M$, $R$, and $T_{\text{eff}}$ values of the individual components (highlighted in dark blue) used to create our binary grid.}
\label{fig: binary_grid}
\end{figure}

We used a subset of the light curves generated using the methodology laid out in Section \ref{subsec: LCsynth} for the creation of our binary grid, ensuring that the subset was relatively representative of the entire grid in terms of $M$, $R$, $T_{\text{eff}}$ and $\nu_{\text{max}}$. This was done in order to keep the binary grid, which comprises combinations of each star with every other star, relatively manageable in size. The subset comprises 30 stars, and their $M$, $R$, and $T_{\text{eff}}$ distributions are displayed in Figure \ref{fig: binary_grid}. 

We therefore generated a total of two sets of 435 (i.e. ${C^{30}_{2}}$) binary light curves, with one set consisting of binaries with only one pulsating component, and the other consisting of binaries with both components pulsating. These were fit using different granulation model configurations (two and four super-Lorentzians), and we perform a differential analysis by comparing the $\nu_{\text{max}}$ values extracted from our binary-RG PSDs ($\nu_{\text{max, binary}}$) with the maximum likelihood estimates of the $\nu_{\text{max}}$ values extracted from our single-RG PSDs ($\nu_{\text{max, single}}$).

\subsection{Results of red-giant binaries with two pulsating components}
\label{subsec: bin2puls}

\begin{figure*}
\centering
\includegraphics[width=\hsize]{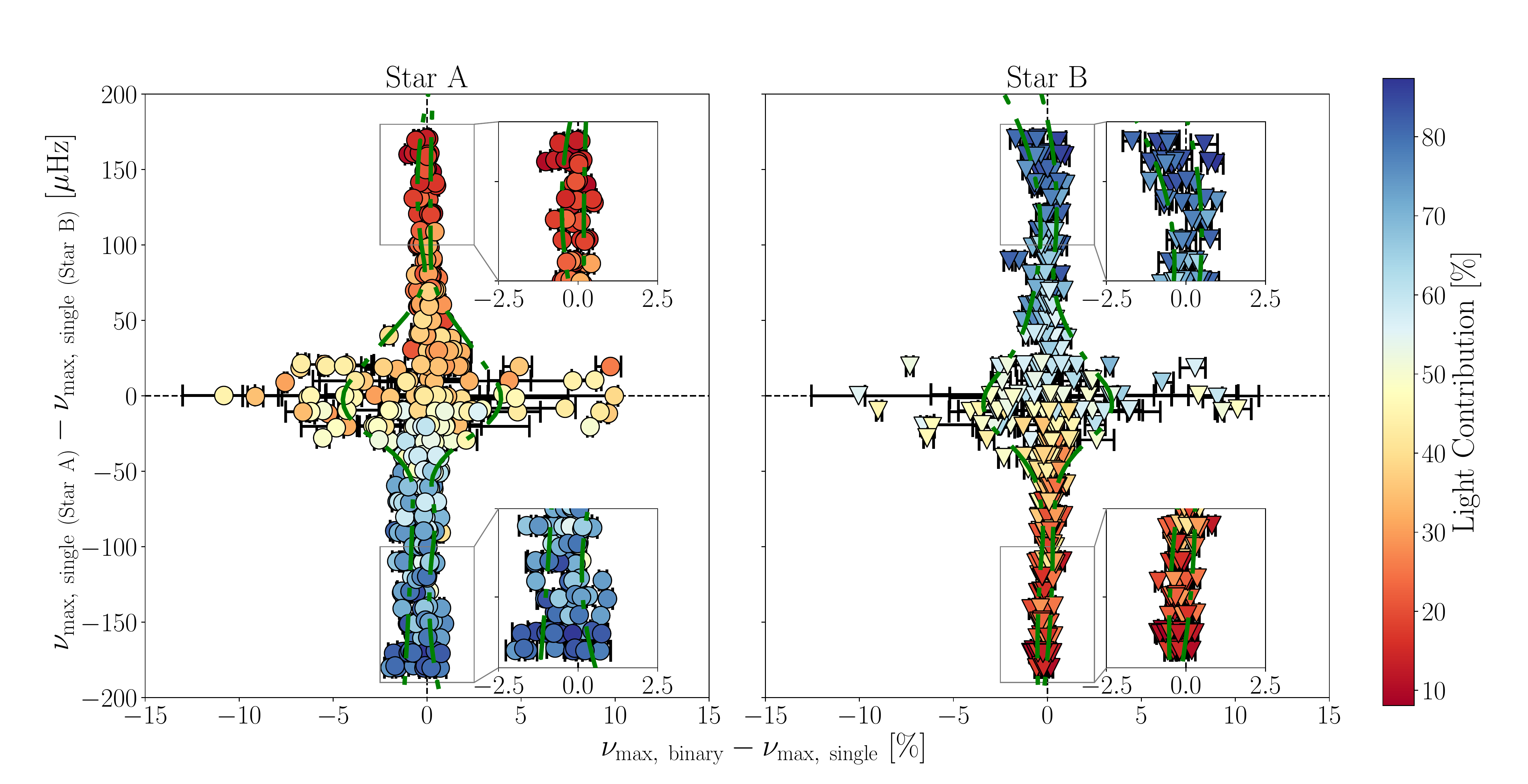}
\caption[A comparison of the $\nu_{\text{max}}$ values of each component (Star A and Star B) extracted from our synthetic RG-binary light curves where both components are pulsating, using a two super-Lorentzian granulation model.]{A comparison of the $\nu_{\text{max}}$ values of each component (Star A, \textit{left panel}; and Star B, \textit{right panel}) extracted from our synthetic RG-binary light curves where both components are pulsating, using a two super-Lorentzian granulation model. The vertical axis corresponds to the difference between the single-RG reference values for Star A ($\nu_{\text{max, single (Star A)}}$) and Star B ($\nu_{\text{max, single (Star B)}}$). The horizontal axis corresponds to the difference between the binary ($\nu_{\text{max, binary}}$) and the single-RG ($\nu_{\text{max, single}}$) reference values. The vertical dashed line represents the zero-point of the difference between $\nu_{\text{max, binary}}$ and $\nu_{\text{max, single}}$. The dash-dotted green lines represent the interpolated $1-\sigma$ level about the mean scatter of the datapoints grouped into 10 $\mu$Hz bins along the vertical axis. The horizontal dashed line represents the zero-point of $\nu_{\text{max, single (Star A)}}-\nu_{\text{max, single (Star B)}}$. The error bars correspond to the 68\% Bayesian credible intervals of the marginalised posterior distributions of the binary fit parameters. The symbols are colour-coded according the percentage light contribution of each component to the binary light curve.}
\label{fig: bin2compgran}

\centering
\includegraphics[width=\hsize]{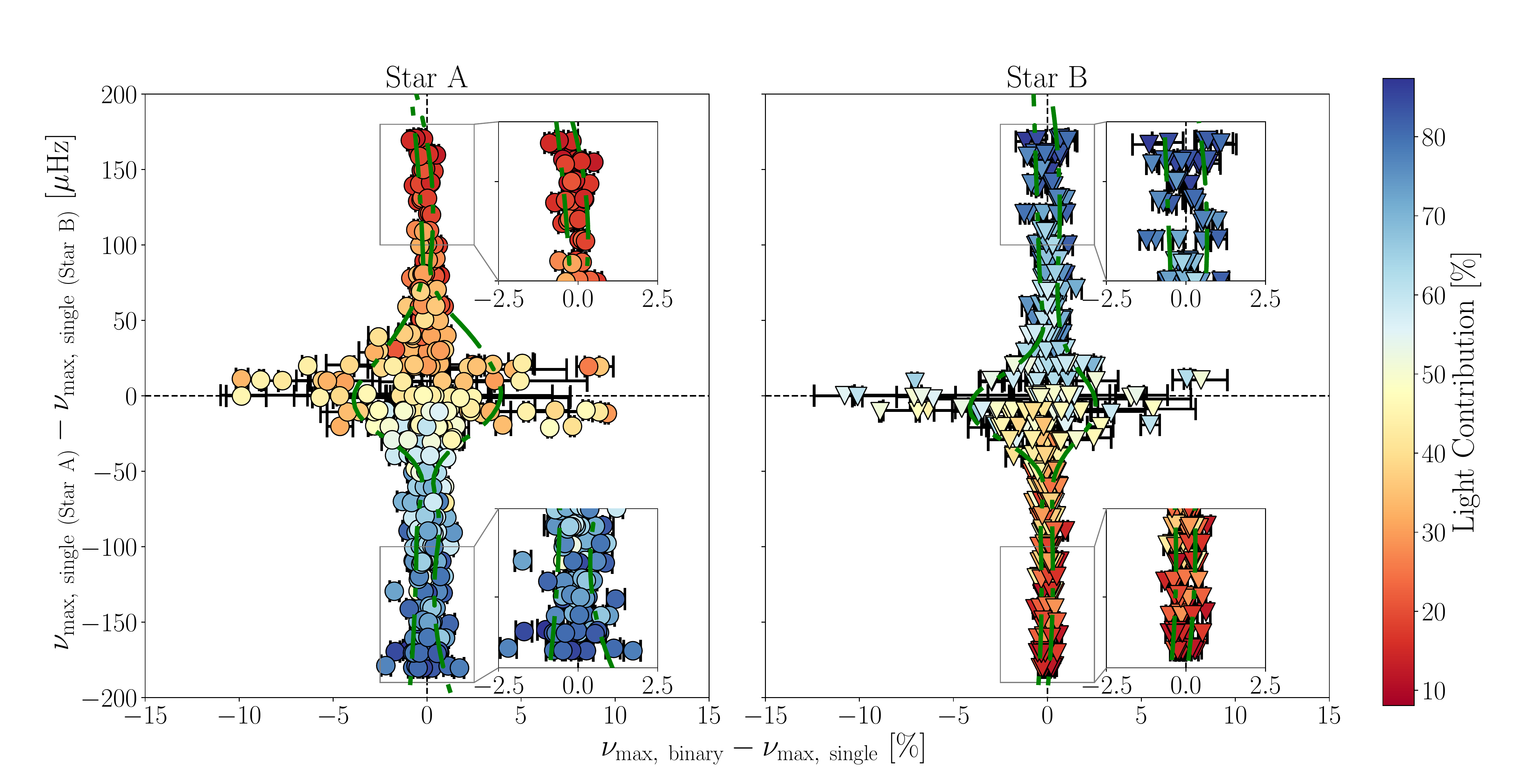}
\caption[A comparison of the $\nu_{\text{max}}$ values of each component (Star A and Star B) extracted from our synthetic RG-binary light curves where both components are pulsating, using a four super-Lorentzian granulation model.]{A comparison of the $\nu_{\text{max}}$ values of each component (Star A, \textit{left panel}; and Star B, \textit{right panel}) extracted from our synthetic RG-binary light curves where both components are pulsating, using a four super-Lorentzian granulation model. The axes, symbols, error bars and colour-coding is the same as in Figure \ref{fig: bin2compgran}.}
\label{fig: bin4compgran}
\end{figure*}

Figure \ref{fig: bin2compgran} and Figure \ref{fig: bin4compgran} show the $\nu_{\text{max}}$ extraction results for a two super-Lorentzian granulation model and a four super-Lorentzian granulation model respectively, where both components (Star A and Star B) in the RG binary are pulsating. The mean percentage difference and the 1-$\sigma$ scatter of the $\nu_{\text{max, binary}}-\nu_{\text{max, single}}$ values, and the mean precision of the $\nu_{\text{max, binary}}$ extraction are detailed in Table \ref{tab: binRG}.

\begin{table*}[!h]
\caption{The mean percentage difference, 1-$\sigma$ scatter and precision of the $\nu_{\text{max, binary}}$ values extracted from our RG-binary PSDs, using either two or four super-Lorentzians to fit the granulation signal.}
\begin{center}
\begin{tabularx}{0.7\linewidth}{YYYY}
 \hline
 \hline
  \multicolumn{4}{c}{$\nu_{\text{max, binary}}-\nu_{\text{max, single}}$ [\%]}\\ 
\hline
\multicolumn{1}{Y}{} & \multicolumn{1}{Y}{Star A} & \multicolumn{1}{Y}{Star B} & \multicolumn{1}{Y}{Single Pulsator}\\
 \hline
\multicolumn{4}{c}{2 Super-Lorentzians}\\
 \hline
 \multicolumn{1}{c}{Mean} & \multicolumn{1}{c}{-0.2} & \multicolumn{1}{c}{-0.06} & \multicolumn{1}{c}{-0.1}\\ 
\multicolumn{1}{c}{1-$\sigma$} & \multicolumn{1}{c}{2.1} & \multicolumn{1}{c}{1.7} & \multicolumn{1}{c}{0.5}\\
\multicolumn{1}{c}{Precision} & \multicolumn{1}{c}{0.8} & \multicolumn{1}{c}{0.7} & \multicolumn{1}{c}{0.3} \\
 \hline
 \multicolumn{4}{c}{4 Super-Lorentzians}\\
 \hline
 \multicolumn{1}{c}{Mean} & \multicolumn{1}{c}{0.008} & \multicolumn{1}{c}{-0.15} & \multicolumn{1}{c}{-0.05}\\ 
\multicolumn{1}{c}{1-$\sigma$} & \multicolumn{1}{c}{1.9} & \multicolumn{1}{c}{1.6} & \multicolumn{1}{c}{0.5}\\
\multicolumn{1}{c}{Precision} & \multicolumn{1}{c}{0.8} & \multicolumn{1}{c}{0.7} & \multicolumn{1}{c}{0.3} \\
 \hline
  \multicolumn{4}{c}{$|\nu_{\mathrm{max, \ binary \ (2SL)}}-\nu_{\mathrm{max, \ binary \ (4SL)}}|$}\\
 \hline
 \multicolumn{1}{c}{Mean} & \multicolumn{1}{c}{1.2} & \multicolumn{1}{c}{0.9} & \multicolumn{1}{c}{0.5}\\ 
\multicolumn{1}{c}{1-$\sigma$} & \multicolumn{1}{c}{2.3} & \multicolumn{1}{c}{1.8} & \multicolumn{1}{c}{0.4}\\
 \hline
\end{tabularx}
\tablefoot{The precision corresponds to the errors propagated from the 68\% Bayesian credible intervals of the marginalised posterior distributions of the fit parameters. The third row of the table showcases the absolute difference in the $\nu_{\text{max}}$ values between the different granulation model configurations (2SL and 4SL).}
\end{center}
\label{tab: binRG}
\end{table*}

The most obvious feature in Figures \ref{fig: bin2compgran} and \ref{fig: bin4compgran} is the large scatter in the central region of the plot, which is where the $\nu_{\text{max}}$ values for Star A ($\nu_{\text{max, single (Star A)}}$) and Star B ($\nu_{\text{max, single (Star B)}}$) are very similar. This scatter is a result of fitting the binary star PSDs where the power excesses of each component overlap (e.g. Figure \ref{fig: binoverlap_fitex}). This necessarily results in degeneracies during the fitting process that are difficult to alleviate without tight constraints on the prior distributions (we use relatively unconstrained priors). In addition, approximately 20\% of the RG/RG binaries with overlapping power excesses displayed non-Gaussian posterior distributions of $\nu_{\text{max}}$ values. This results in the asymmetric, and often large, error bars for the datapoints in the region around the zero-point value of $\nu_{\text{max, single (Star A)}}-\nu_{\text{max, single (Star B)}}$. 

We also observe a marked increase in the scatter of $\nu_{\text{max, binary}}-\nu_{\text{max, single}}$ as the light contribution increases (see zoomed insets in Figure \ref{fig: bin2compgran}). This is once again a consequence of representing our results in percentages, and is also observed in the single-RG case (see Section \ref{sec: SingleRGs}): The component with the higher light contribution tends to have the lower $\nu_{\text{max}}$ (as $\nu_{\text{max}}$ is inversely proportional to $L$), and therefore similar absolute $\nu_{\text{max, binary}}-\nu_{\text{max, single}}$ values are larger in terms of percentage at lower $\nu_{\text{max, single}}$ values.

Our results shown in Table \ref{tab: binRG} indicate that there is no appreciable systematic offset in our $\nu_{\text{max, binary}}$ values when compared to our $\nu_{\text{max, single}}$ values. The small residual systematic offsets that remain are certainly well below the 1-$\sigma$ scatter reported in the single-RG case (cf. Table \ref{tab: SingleRG}). However, the 1-$\sigma$ scatter of the $\nu_{\text{max, binary}}$ values is much higher, and the mean precision is much lower than in the single-star case. It must be noted that these results represent the global properties of the entire grid, and are therefore largely biased by the datapoints corresponding to overlapping power excesses, where a large 1-$\sigma$ scatter and low precision are expected. For the cases where the power excesses are well separated (e.g. Figure \ref{fig: binseparated_fitex}), we clearly see a much smaller scatter that is quantitatively comparable to the single star case discussed in the Section \ref{sec: SingleRGs}.

\begin{figure}[t]
\centering
\includegraphics[width=\hsize]{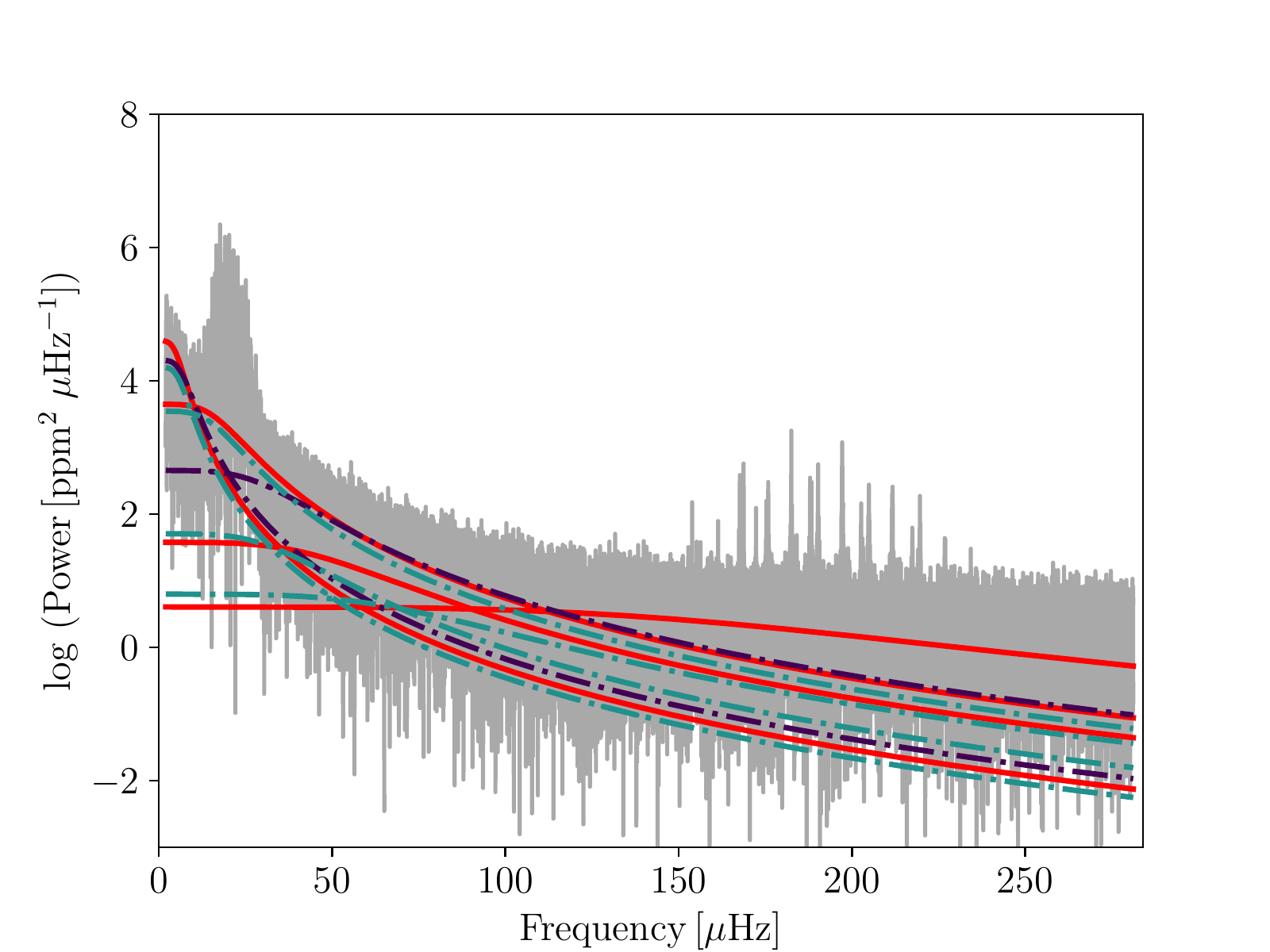}
\caption[A comparison of the input granulation signal with the best-fit.]{A comparison of the input granulation signal with those from the maximum-likelihood values of the two and four super-Lorentzian granulation model fits, for one of our binary-RG PSDs where both components are pulsating. The super-Lorentzians used as inputs are represented by the solid red curves, with the two and four super-Lorentzians fits represented by the dark blue and blue-green dash-dotted lines respectively.}
\label{fig: gran_degen_2puls}
\end{figure}

Another interesting observation is that there is little difference in the overall results of the two and four super-Lorentzian component fits. This can be attributed once again to the degenerate nature of the background fit: adding additional super-Lorentzians does not seem to improve the quality of the fit. This is in agreement with the conclusions of \cite{Kallinger2014} and is demonstrated in Figure \ref{fig: gran_degen_2puls}, which shows the input super-Lorentzians for one of our synthetic PSDs compared with those derived from the maximum-likelihood values for the two and four super-Lorentzian component granulation models. The background fit is so degenerate that i) two super-Lorentzians can still adequately model the background (verifying the approach of \citealt{Beck2018a}), and ii) the input parameters are not recovered using the four super-Lorentzian model.

Overall, we observe an intrinsic scatter of the order of 2\% in $\nu_{\text{max}}$, corresponding to a scatter of 6\% in $M$ and 6\% in $R$. We once again stress that the larger scatter in $\nu_{\text{max}}$ is a direct consequence of the additional fitting degeneracy under the condition of the overlapping power excesses of the two pulsating components: Our uncertainties are otherwise comparable to the single-RG case and do not exceed the typical uncertainties for the scaling relations reported in the literature (cf. Section \ref{sec: SingleRGs}).

\begin{figure*}
\centering
\includegraphics[width=\hsize]{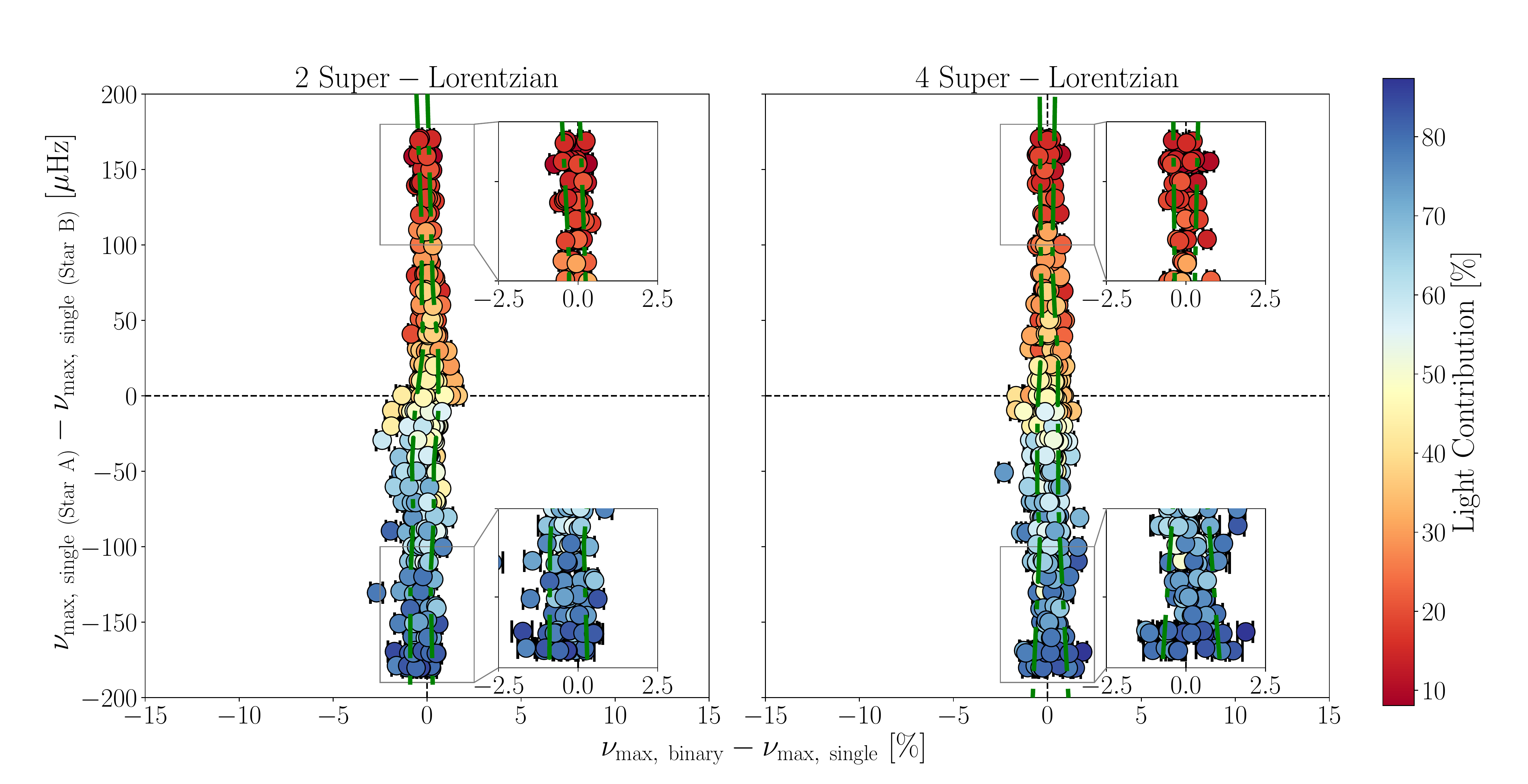}
\caption[A comparison of the $\nu_{\text{max}}$ values for a two super-Lorentzian (\textit{left panel}) and a four super-Lorentzian (\textit{right panel}) granulation model, extracted from our synthetic RG-binary light curves where only one component (Star A) is pulsating.]{A comparison of the $\nu_{\text{max}}$ values for a two super-Lorentzian (\textit{left panel}) and a four super-Lorentzian (\textit{right panel}) granulation model, extracted from our synthetic RG-binary light curves where only one component (Star A) is pulsating. The axes, symbols, error bars and colour-coding is the same as in Figure \ref{fig: bin2compgran}.}
\label{fig: bin2comp4comp1puls}
\end{figure*}

\subsection{Results of red-giant binaries with one pulsating component}
\label{subsec: bin1puls}

Figure \ref{fig: bin2comp4comp1puls} shows the $\nu_{\text{max}}$ extraction results for the two and four super-Lorentzian component granulation models, where only one of the two components (Star A) in the RG binary is pulsating (e.g. Figure \ref{fig: bin1puls_fitex}). The mean percentage difference and the 1-$\sigma$ scatter of the $\nu_{\text{max, binary}}-\nu_{\text{max, single}}$ values, and the mean precision of the $\nu_{\text{max, binary}}$ extraction are also detailed in the last column of Table \ref{tab: binRG}.

Once again, our results show that there is no appreciable systematic offset in our $\nu_{\text{max, binary}}$ values when compared to our $\nu_{\text{max, single}}$ values. In fact, the results that we had obtained were very similar to that of the single-RG case: The 1-$\sigma$ scatter and precision of the $\nu_{\text{max}}$ extraction is almost identical to that of the single-RG case (detailed in Table \ref{tab: SingleRG}), with perhaps a slight decrease in mean precision (0.3\% vs. 0.2\%). We do not observe the large 1-$\sigma$ scatter that was observed in the double-pulsator case, as was expected due to the absence of a second power excess that would add significant degeneracy in the overlapping case. We do, however, observe the increase in the scatter of $\nu_{\text{max, binary}}-\nu_{\text{max, single}}$ as the light contribution increases (see zoomed insets in Figure \ref{fig: bin2comp4comp1puls}) as per the double-pulsator case, which is, as mentioned, due to the representation of our results in percentages.

\begin{figure}[t]
\centering
\includegraphics[width=\hsize]{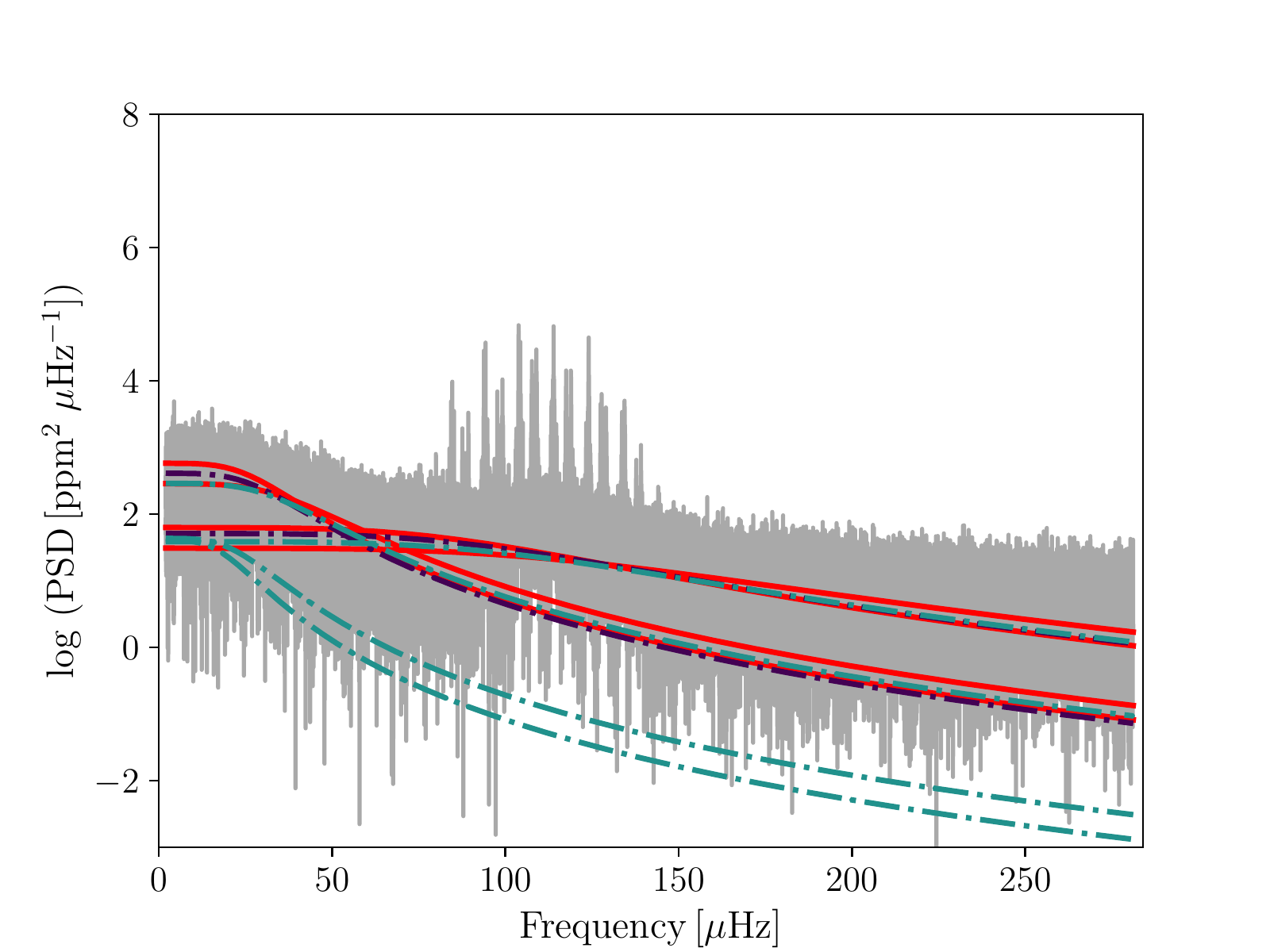}
\caption[A comparison of the input granulation signal with the best-fit.]{A comparison of the input granulation signal with those from the maximum-likelihood values of the two and four super-Lorentzian granulation model fits, for one of our binary-RG PSDs where only one component is pulsating. The super-Lorentzians used as inputs are represented by the solid red curves, with the two and four super-Lorentzian fits represented by the dark blue and blue-green dash-dotted lines respectively.}
\label{fig: gran_degen_1puls}
\end{figure}

Similar to our binary-RG results, we once again observe very small differences between the different granulation model configurations: a consequence of the degeneracy of the background fit. This is demonstrated in Figure \ref{fig: gran_degen_1puls}, which shows the input super-Lorentzians for one of our single-pulsator synthetic PSDs compared with those derived from the maximum-likelihood values for the two and four super-Lorentzian component granulation models. Once again, we observe that the background fit is so degenerate that i) two super-Lorentzians can still adequately model the background, and ii) the input parameters are not recovered using the four super-Lorentzian model. This is in line with the conclusions of \cite{Kallinger2014} that there is a limit to the number of granulation components that can be uniquely modelled for a sufficiently-bright ($K_{\text{p}}<12$) RG.

Overall, we observe an intrinsic scatter of the order of 0.5\% in $\nu_{\text{max}}$, corresponding to a scatter of 1.5\% in $M$ and 0.5\% in $R$, which is compatible with the single-RG case. It should be noted that the uncertainties typically reported for the scaling relations are larger than, or of the order of, the combined errors reported here. 

\subsection{Comparison between the different granulation model configurations}
\label{subsec: granmodelcomp}

\begin{figure*}
\centering
\includegraphics[width=\hsize]{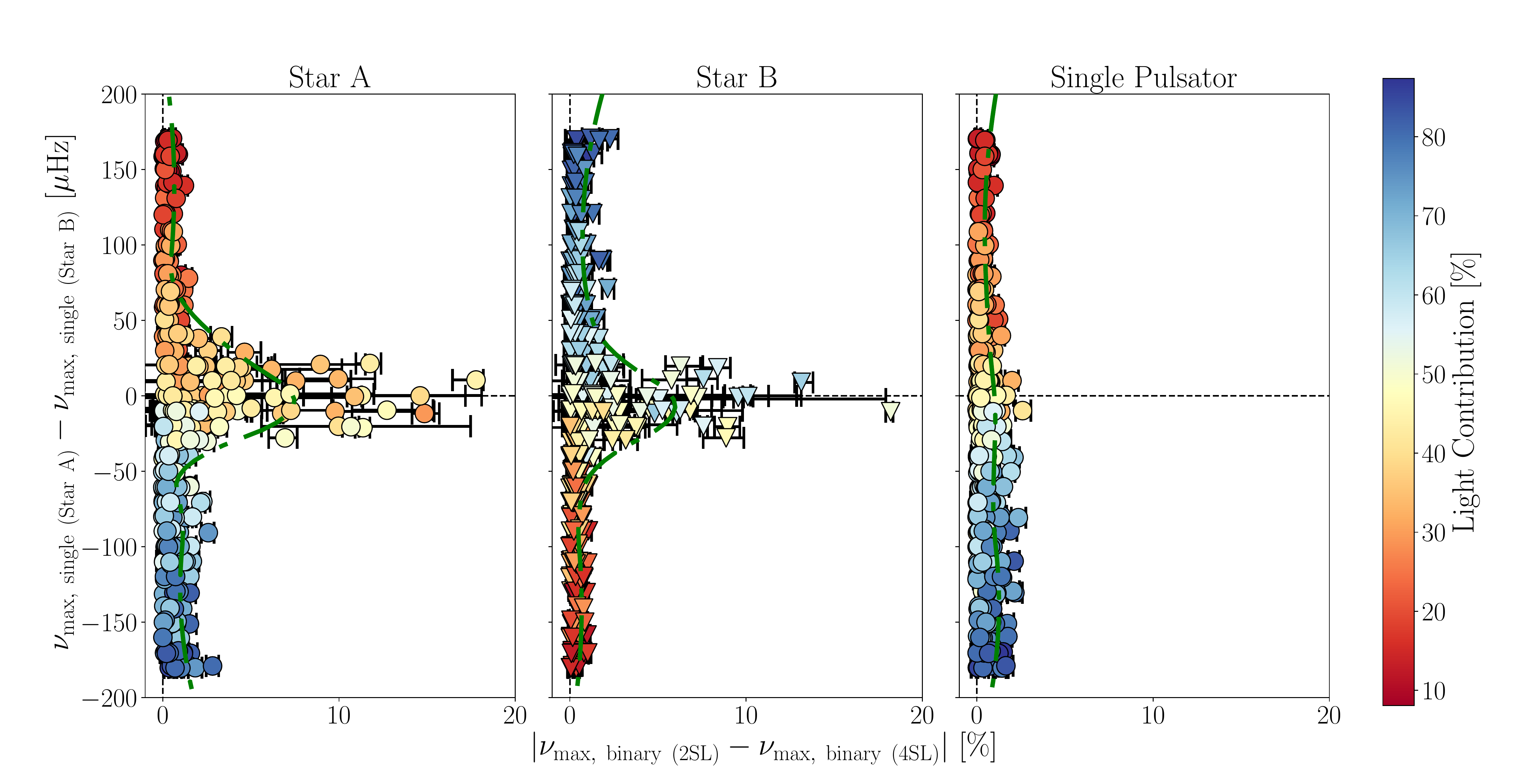}
\caption[The difference in the $\nu_{\text{max}}$ values extracted from the fits using a two super-Lorentzian and a four super-Lorentzian granulation model.]{The difference in the $\nu_{\text{max}}$ values extracted from the fits using a two super-Lorentzian and a four super-Lorentzian granulation model. \textit{Left and middle panels}: The difference in $\nu_{\text{max, fit}}$ for the the RG-binary light curves where both components (Star A and Star B) are pulsating. \textit{Right panel}: The difference in $\nu_{\text{max, fit}}$ for the RG-binary light curves where only one component (Star A) is pulsating. The axes, symbols, error bars and colour-coding is the same as in Figure \ref{fig: bin2compgran}, except that the horizontal axis corresponds to the absolute percentage difference between the $\nu_{\text{max, fit}}$ values for the two super-Lorentzian $\text{(2SL)}$ and four super-Lorentzian $\text{(4SL)}$ cases.}
\label{fig: bin2comp4compdiff}
\end{figure*}

We performed a comparison of the two different granulation model configurations by taking the absolute difference of $\nu_{\text{max, binary}}$ values extracted from the fits using each type of granulation model, for light curves with two pulsating components and a single pulsating component. The left and middle panels of Figure \ref{fig: bin2comp4compdiff} show the difference in $\nu_{\text{max, binary}}$ for the the RG-binary light curves where both components (Star A and Star B) are pulsating, and the right panel shows the difference in $\nu_{\text{max, binary}}$ for the RG-binary light curves where only one component (Star A) is pulsating. All of these results are detailed in the bottom row of Table \ref{tab: binRG}, where we show the mean percentage difference and the 1-$\sigma$ scatter of the $|\nu_{\text{max, binary (2SL)}}-\nu_{\text{max, binary (4SL)}}|$ values.

Once again, the greatest percentage difference between the $\nu_{\text{max, fit}}$ values extracted from fits using the two and four super-Lorentzian granulation models is in the region around the zero-point value of $\nu_{\text{max, single (Star A)}}-\nu_{\text{max, single (Star B)}}$. As mentioned, the $\nu_{\text{max}}$ extraction from PSDs with overlapping power excesses is generally unreliable, and the different granulation models function as an additional confounding factor in these cases. It can also be seen that the differences between the different granulation model configurations increases as with increasing light contribution, which is once again a consequence of expressing our results in percentages.

In general, it can be seen that the difference between the $\nu_{\text{max}}$ values for the two and four super-Lorentzian granulation model configurations are small outside of the case of overlapping power excesses, with differences and scatter of the order of 0.5\% in $\nu_{\text{max}}$, corresponding to 1.5\% in $M$ and 0.5\% in $R$.

\section{Discussion and conclusions}
\label{sec: conclusions}

This paper details the results of our study, where we attempted to estimate the impact of a contaminating photometric signal with a distinct background profile on the extraction of the global asteroseismic parameter $\nu_{\text{max}}$. We developed a robust methodology for the simulation of the light curves of first-ascent RGB single- and binary stars, incorporating frequency-dependent modewidths and luminosity-dependent white noise.   

We simulated the light curves of single RGs, and RG/RG binaries with i) one pulsating component and ii) two pulsating components and extracted the $\nu_{\text{max}}$ values of the pulsating component. We also tested two different granulation model configurations for our RG binaries: i) a two super-Lorentzian and ii) a four super-Lorentzian granulation profile and computed the differences between the binary-RG and single-RG reference $\nu_{\text{max}}$ values using each type of granulation model configuration. A summary of these results is presented below:

\begin{itemize}
\item {Single RGs}
\begin{itemize}
\item {Systematics: We find a systematic underestimation of 1\% in $\nu_{\text{max}}$, 2.5\% in $M$ and 1\% in $R$. This is caused by asymmetric mode distribution in the power excesses (see Figure \ref{fig: PulsInp}), resulting in differences between the $\nu_{\text{max}}$ values derived from the fit and from scaling relations.}
\item {Granulation: We confirm the randomising effect of the granulation signal, }slightly increasing the degree of systematic underestimation, the intrinsic scatter, and the precision of extraction. We also find that the $\nu_{\text{max}}$ values extracted from our single-RG PSDs with and without granulation often did not showcase agreement within the formal errors.
\item {White Noise: We find that white noise has a negligible effect on the extraction of $\nu_{\text{max}}$ for stars with $K_{\text{p}}<12$, as noted by \cite{Kallinger2014}.}
\end{itemize}
\item {Binary RGs}
\begin{itemize}
\item {Systematics: We find negligible systematic differences between the binary-RG and single-RG $\nu_{\text{max}}$ values, confirming that the systematic offset observed in the single-RG case is a methodological issue.}
\item {Two pulsating components: We find that the $\nu_{\text{max}}$ extraction for PSDs featuring overlapping power excesses was highly unreliable, as was expected due to the use of unconstrained Gaussians to fit the power excesses. Outside of this scenario, we find that our results are very similar to the single-RG case.}
\item {One pulsating component: We find that our results are very similar to the single-RG case, showing that the light contribution of the individual components has minimal effect on the $\nu_{\text{max}}$ extraction for cases where the light contribution of the component is above $\sim$ 10\%.}
\item {Granulation model configurations: We find minimal difference in the results when using either two or four super-Lorentzian components to represent the granulation signal. This is due to the degeneracy of the granulation fits such that both model configurations are able to represent it equally well.}
\end{itemize}
\end{itemize}

Our results indicate that binarity has a strong effect on the extraction of $\nu_{\text{max}}$ in configurations where both components are pulsating and their power excesses overlap in the frequency domain. These particular binary configurations need a special treatment when using binarity for testing and validating scaling relations (e.g. such as using methodology proposed by \citealt{Beck2018a}). Outside of the abovementioned scenario, irrespective of individual light contributions of the two stars, the extraction of $\nu_{\text{max}}$ is not affected by binarity and is only subject to systematic uncertainties due to the adopted methodology. 

Our findings indicate that a systematic offset of observationally-derived parameters from scaling-relation derived parameters might be a result of the discrepancy between the observed $\nu_{\text{max}}$, which we found to be highly sensitive to mode distribution in the power excess, and the theoretical $\nu_{\text{max}}$. Our results also indicate that the $\sim$15\%/5\% overestimation of seismic versus dynamical mass/radius reported by \cite{Gaulme2016} is unrelated to the effect of photometric contamination due to binarity and hence requires alternative explanations. We also find that while granulation does function as a randomisation element, resulting in discrepancies in the extraction of $\nu_{\text{max}}$, the differences in $\nu_{\text{max}}$ extraction between different methods (as reported in \citealt{Hekker2011a}) are significantly larger than the granulation discrepancy. In addition, the minimal difference in our binary results when using either two or four super-Lorentzian components to represent the granulation indicates that the current physical interpretation of the granulation terms is suspect. Additional investigation of the physics of granulation and its observational signature would be required before the degeneracy in the granulation fit can be lifted.

Our results indicate that the presence of systematic errors in the extraction of $\nu_{\text{max}}$ are a consequence of the fitting methodology. We find that while these uncertainties are smaller than or of the order of typical uncertainties reported for the scaling relations in the literature, they still worth considering when interpreting results in the context of observed discrepancies between scaling relations-based masses/radii and those inferred from binary dynamics. We can therefore conclude that photometric contamination, besides decreasing the signal-to-noise ratios of the individual components in the binary PSD, would have a negligible effect on the extraction of the global asteroseismic parameter $\nu_{\text{max}}$ from \textsc{tess} data.

\begin{acknowledgements}

The authors would like to thank the anonymous referee for the constructive criticism that had radically transformed the manuscript into the form seen here. The research leading to these results has received funding from the Fonds Wetenschappelijk Onderzoek - Vlaanderen (FWO) under the grant agreement G0H5416N (ERC Opvangproject), and from the European Research Council (ERC) under the European Union’s Horizon 2020 research and innovation programme (grant agreement no. 670519: MAMSIE). PGB acknowledges the support of the MINECO under the program 'Juan de la Cierva incorporacion' (IJCI-2015-26034). The authors would also like to thank Prof. C. Aerts and the MAMSIE team for useful discussions.

\end{acknowledgements}

\bibliographystyle{aa}
\bibliography{RG}

\newcommand{\noop}[1]{}
\begin{thebibliography}{70}
\expandafter\ifx\csname natexlab\endcsname\relax\def\natexlab#1{#1}\fi

\bibitem[{{Appourchaux} {et~al.}(2015){Appourchaux}, {Antia}, {Ball},
  {Creevey}, {Lebreton}, {Verma}, {Vorontsov}, {Campante}, {Davies}, {Gaulme},
  {R{\'e}gulo}, {Horch}, {Howell}, {Everett}, {Ciardi}, {Fossati}, {Miglio},
  {Montalb{\'a}n}, {Chaplin}, {Garc{\'{\i}}a}, \& {Gizon}}]{Appourchaux2015}
{Appourchaux}, T., {Antia}, H.~M., {Ball}, W., {et~al.} 2015, \aap, 582, A25

\bibitem[{{Appourchaux} {et~al.}(2014){Appourchaux}, {Antia}, {Benomar},
  {Campante}, {Davies}, {Handberg}, {Howe}, {R{\'e}gulo}, {Belkacem}, {Houdek},
  {Garc{\'{\i}}a}, \& {Chaplin}}]{Appourchaux2014}
{Appourchaux}, T., {Antia}, H.~M., {Benomar}, O., {et~al.} 2014, \aap, 566, A20

\bibitem[{{Arentoft} {et~al.}(2008){Arentoft}, {Kjeldsen}, {Bedding}, {Bazot},
  {Christensen-Dalsgaard}, {Dall}, {Karoff}, {Carrier}, {Eggenberger},
  {Sosnowska}, {Wittenmyer}, {Endl}, {Metcalfe}, {Hekker}, {Reffert}, {Butler},
  {Bruntt}, {Kiss}, {O'Toole}, {Kambe}, {Ando}, {Izumiura}, {Sato}, {Hartmann},
  {Hatzes}, {Bouchy}, {Mosser}, {Appourchaux}, {Barban}, {Berthomieu},
  {Garcia}, {Michel}, {Provost}, {Turck-Chi{\`e}ze}, {Marti{\'c}}, {Lebrun},
  {Schmitt}, {Bertaux}, {Bonanno}, {Benatti}, {Claudi}, {Cosentino}, {Leccia},
  {Frandsen}, {Brogaard}, {Glowienka}, {Grundahl}, \&
  {Stempels}}]{Arentoft2008}
{Arentoft}, T., {Kjeldsen}, H., {Bedding}, T.~R., {et~al.} 2008, \apj, 687,
  1180

\bibitem[{{Asplund} {et~al.}(2009){Asplund}, {Grevesse}, {Sauval}, \&
  {Scott}}]{Asplund2009}
{Asplund}, M., {Grevesse}, N., {Sauval}, A.~J., \& {Scott}, P. 2009, \araa, 47,
  481

\bibitem[{{Auvergne} {et~al.}(2009){Auvergne}, {Bodin}, {Boisnard}, {Buey},
  {Chaintreuil}, {Epstein}, {Jouret}, {Lam-Trong}, {Levacher}, {Magnan},
  {Perez}, {Plasson}, {Plesseria}, {Peter}, {Steller}, {Tiph{\`e}ne}, {Baglin},
  {Agogu{\'e}}, {Appourchaux}, {Barbet}, {Beaufort}, {Bellenger}, {Berlin},
  {Bernardi}, {Blouin}, {Boumier}, {Bonneau}, {Briet}, {Butler}, {Cautain},
  {Chiavassa}, {Costes}, {Cuvilho}, {Cunha-Parro}, {de Oliveira Fialho},
  {Decaudin}, {Defise}, {Djalal}, {Docclo}, {Drummond}, {Dupuis}, {Exil},
  {Faur{\'e}}, {Gaboriaud}, {Gamet}, {Gavalda}, {Grolleau}, {Gueguen},
  {Guivarc'h}, {Guterman}, {Hasiba}, {Huntzinger}, {Hustaix}, {Imbert},
  {Jeanville}, {Johlander}, {Jorda}, {Journoud}, {Karioty}, {Kerjean},
  {Lafond}, {Lapeyrere}, {Landiech}, {Larqu{\'e}}, {Laudet}, {Le Merrer},
  {Leporati}, {Leruyet}, {Levieuge}, {Llebaria}, {Martin}, {Mazy}, {Mesnager},
  {Michel}, {Moalic}, {Monjoin}, {Naudet}, {Neukirchner}, {Nguyen-Kim},
  {Ollivier}, {Orcesi}, {Ottacher}, {Oulali}, {Parisot}, {Perruchot},
  {Piacentino}, {Pinheiro da Silva}, {Platzer}, {Pontet}, {Pradines},
  {Quentin}, {Rohbeck}, {Rolland}, {Rollenhagen}, {Romagnan}, {Russ}, {Samadi},
  {Schmidt}, {Schwartz}, {Sebbag}, {Smit}, {Sunter}, {Tello}, {Toulouse},
  {Ulmer}, {Vandermarcq}, {Vergnault}, {Wallner}, {Waultier}, \&
  {Zanatta}}]{Auvergne2009}
{Auvergne}, M., {Bodin}, P., {Boisnard}, L., {et~al.} 2009, \aap, 506, 411

\bibitem[{{Beck} {et~al.}(2014){Beck}, {Hambleton}, {Vos}, {Kallinger},
  {Bloemen}, {Tkachenko}, {Garc{\'{\i}}a}, {{\O}stensen}, {Aerts}, {Kurtz}, {De
  Ridder}, {Hekker}, {Pavlovski}, {Mathur}, {De Smedt}, {Derekas}, {Corsaro},
  {Mosser}, {Van Winckel}, {Huber}, {Degroote}, {Davies}, {Pr{\v s}a},
  {Debosscher}, {Elsworth}, {Nemeth}, {Siess}, {Schmid}, {P{\'a}pics}, {de
  Vries}, {van Marle}, {Marcos-Arenal}, \& {Lobel}}]{Beck2014}
{Beck}, P.~G., {Hambleton}, K., {Vos}, J., {et~al.} 2014, \aap, 564, A36

\bibitem[{{Beck} {et~al.}(2018{\natexlab{a}}){Beck}, {Kallinger}, {Pavlovski},
  {Palacios}, {Tkachenko}, {Mathis}, {Garc{\'\i}a}, {Corsaro}, {Johnston},
  {Mosser}, {Ceillier}, {do Nascimento}, \& {Raskin}}]{Beck2018a}
{Beck}, P.~G., {Kallinger}, T., {Pavlovski}, K., {et~al.} 2018{\natexlab{a}},
  \aap, 612, A22

\bibitem[{{Beck} {et~al.}(2018{\natexlab{b}}){Beck}, {Mathis}, {Gallet},
  {Charbonnel}, {Benbakoura}, {Garc{\'{\i}}a}, \& {do Nascimento}}]{Beck2018b}
{Beck}, P.~G., {Mathis}, S., {Gallet}, F., {et~al.} 2018{\natexlab{b}}, \mnras,
  479, L123

\bibitem[{{Belkacem} {et~al.}(2011){Belkacem}, {Goupil}, {Dupret}, {Samadi},
  {Baudin}, {Noels}, \& {Mosser}}]{Belkacem2011}
{Belkacem}, K., {Goupil}, M.~J., {Dupret}, M.~A., {et~al.} 2011, \aap, 530,
  A142

\bibitem[{{Belkacem} {et~al.}(2013){Belkacem}, {Samadi}, {Mosser}, {Goupil}, \&
  {Ludwig}}]{Belkacem2013}
{Belkacem}, K., {Samadi}, R., {Mosser}, B., {Goupil}, M.-J., \& {Ludwig}, H.-G.
  2013, in Astronomical Society of the Pacific Conference Series, Vol. 479,
  Progress in Physics of the Sun and Stars: A New Era in Helio- and
  Asteroseismology, ed. H.~{Shibahashi} \& A.~E. {Lynas-Gray}, 61

\bibitem[{{Bellinger} {et~al.}(2017){Bellinger}, {Basu}, {Hekker}, \&
  {Ball}}]{Bellinger2017}
{Bellinger}, E.~P., {Basu}, S., {Hekker}, S., \& {Ball}, W.~H. 2017, \apj, 851,
  80

\bibitem[{{Borucki} {et~al.}(2010){Borucki}, {Koch}, {Basri}, {Batalha},
  {Brown}, {Caldwell}, {Caldwell}, {Christensen-Dalsgaard}, {Cochran},
  {DeVore}, {Dunham}, {Dupree}, {Gautier}, {Geary}, {Gilliland}, {Gould},
  {Howell}, {Jenkins}, {Kondo}, {Latham}, {Marcy}, {Meibom}, {Kjeldsen},
  {Lissauer}, {Monet}, {Morrison}, {Sasselov}, {Tarter}, {Boss}, {Brownlee},
  {Owen}, {Buzasi}, {Charbonneau}, {Doyle}, {Fortney}, {Ford}, {Holman},
  {Seager}, {Steffen}, {Welsh}, {Rowe}, {Anderson}, {Buchhave}, {Ciardi},
  {Walkowicz}, {Sherry}, {Horch}, {Isaacson}, {Everett}, {Fischer}, {Torres},
  {Johnson}, {Endl}, {MacQueen}, {Bryson}, {Dotson}, {Haas}, {Kolodziejczak},
  {Van Cleve}, {Chandrasekaran}, {Twicken}, {Quintana}, {Clarke}, {Allen},
  {Li}, {Wu}, {Tenenbaum}, {Verner}, {Bruhweiler}, {Barnes}, \&
  {Prsa}}]{Borucki2010}
{Borucki}, W.~J., {Koch}, D., {Basri}, G., {et~al.} 2010, Science, 327, 977

\bibitem[{{Brogaard} {et~al.}(2018){Brogaard}, {Hansen}, {Miglio}, {Slumstrup},
  {Frandsen}, {Jessen-Hansen}, {Lund}, {Bossini}, {Thygesen}, {Davies},
  {Chaplin}, {Arentoft}, {Bruntt}, {Grundahl}, \& {Handberg}}]{Brogaard2018}
{Brogaard}, K., {Hansen}, C.~J., {Miglio}, A., {et~al.} 2018, \mnras, 476, 3729

\bibitem[{{Brogaard} {et~al.}(2016){Brogaard}, {Jessen-Hansen}, {Handberg},
  {Arentoft}, {Frandsen}, {Grundahl}, {Bruntt}, {Sandquist}, {Miglio}, {Beck},
  {Thygesen}, {Kj{\ae}rgaard}, \& {Haugaard}}]{Brogaard2016}
{Brogaard}, K., {Jessen-Hansen}, J., {Handberg}, R., {et~al.} 2016,
  Astronomische Nachrichten, 337, 793

\bibitem[{{Brown} {et~al.}(1991){Brown}, {Gilliland}, {Noyes}, \&
  {Ramsey}}]{Brown1991}
{Brown}, T.~M., {Gilliland}, R.~L., {Noyes}, R.~W., \& {Ramsey}, L.~W. 1991,
  \apj, 368, 599

\bibitem[{{Chaplin} {et~al.}(2014){Chaplin}, {Basu}, {Huber}, {Serenelli},
  {Casagrande}, {Silva Aguirre}, {Ball}, {Creevey}, {Gizon}, {Handberg},
  {Karoff}, {Lutz}, {Marques}, {Miglio}, {Stello}, {Suran}, {Pricopi},
  {Metcalfe}, {Monteiro}, {Molenda-{\.Z}akowicz}, {Appourchaux},
  {Christensen-Dalsgaard}, {Elsworth}, {Garc{\'{\i}}a}, {Houdek}, {Kjeldsen},
  {Bonanno}, {Campante}, {Corsaro}, {Gaulme}, {Hekker}, {Mathur}, {Mosser},
  {R{\'e}gulo}, \& {Salabert}}]{Chaplin2014}
{Chaplin}, W.~J., {Basu}, S., {Huber}, D., {et~al.} 2014, \apjs, 210, 1

\bibitem[{{Chaplin} {et~al.}(2011){Chaplin}, {Kjeldsen},
  {Christensen-Dalsgaard}, {Basu}, {Miglio}, {Appourchaux}, {Bedding},
  {Elsworth}, {Garc{\'{\i}}a}, {Gilliland}, {Girardi}, {Houdek}, {Karoff},
  {Kawaler}, {Metcalfe}, {Molenda-{\.Z}akowicz}, {Monteiro}, {Thompson},
  {Verner}, {Ballot}, {Bonanno}, {Brand{\~a}o}, {Broomhall}, {Bruntt},
  {Campante}, {Corsaro}, {Creevey}, {Do{\u g}an}, {Esch}, {Gai}, {Gaulme},
  {Hale}, {Handberg}, {Hekker}, {Huber}, {Jim{\'e}nez}, {Mathur}, {Mazumdar},
  {Mosser}, {New}, {Pinsonneault}, {Pricopi}, {Quirion}, {R{\'e}gulo},
  {Salabert}, {Serenelli}, {Silva Aguirre}, {Sousa}, {Stello}, {Stevens},
  {Suran}, {Uytterhoeven}, {White}, {Borucki}, {Brown}, {Jenkins}, {Kinemuchi},
  {Van Cleve}, \& {Klaus}}]{Chaplin2011}
{Chaplin}, W.~J., {Kjeldsen}, H., {Christensen-Dalsgaard}, J., {et~al.} 2011,
  Science, 332, 213

\bibitem[{{Corsaro} {et~al.}(2013){Corsaro}, {Fr{\"o}hlich}, {Bonanno},
  {Huber}, {Bedding}, {Benomar}, {De Ridder}, \& {Stello}}]{Corsaro2013}
{Corsaro}, E., {Fr{\"o}hlich}, H.-E., {Bonanno}, A., {et~al.} 2013, \mnras,
  430, 2313

\bibitem[{{De Ridder} {et~al.}(2006){De Ridder}, {Arentoft}, \&
  {Kjeldsen}}]{DeRidder2006}
{De Ridder}, J., {Arentoft}, T., \& {Kjeldsen}, H. 2006, \mnras, 365, 595

\bibitem[{{De Ridder} {et~al.}(2009){De Ridder}, {Barban}, {Baudin}, {Carrier},
  {Hatzes}, {Hekker}, {Kallinger}, {Weiss}, {Baglin}, {Auvergne}, {Samadi},
  {Barge}, \& {Deleuil}}]{DeRidder2009}
{De Ridder}, J., {Barban}, C., {Baudin}, F., {et~al.} 2009, \nat, 459, 398

\bibitem[{{De Ridder} {et~al.}(2016){De Ridder}, {Molenberghs}, {Eyer}, \&
  {Aerts}}]{DeRidder2016}
{De Ridder}, J., {Molenberghs}, G., {Eyer}, L., \& {Aerts}, C. 2016, \aap, 595,
  L3

\bibitem[{{Dupret} {et~al.}(2009){Dupret}, {Belkacem}, {Samadi}, {Montalban},
  {Moreira}, {Miglio}, {Godart}, {Ventura}, {Ludwig}, {Grigahc{\`e}ne},
  {Goupil}, {Noels}, \& {Caffau}}]{Dupret2009}
{Dupret}, M.-A., {Belkacem}, K., {Samadi}, R., {et~al.} 2009, \aap, 506, 57

\bibitem[{{Foreman-Mackey} {et~al.}(2013){Foreman-Mackey}, {Hogg}, {Lang}, \&
  {Goodman}}]{ForemanMackey2013}
{Foreman-Mackey}, D., {Hogg}, D.~W., {Lang}, D., \& {Goodman}, J. 2013, \pasp,
  125, 306

\bibitem[{{Frandsen} {et~al.}(2013){Frandsen}, {Lehmann}, {Hekker},
  {Southworth}, {Debosscher}, {Beck}, {Hartmann}, {Pigulski}, {Kopacki},
  {Ko{\l}aczkowski}, {St{\c e}{\'s}licki}, {Thygesen}, {Brogaard}, \&
  {Elsworth}}]{Frandsen2013}
{Frandsen}, S., {Lehmann}, H., {Hekker}, S., {et~al.} 2013, \aap, 556, A138

\bibitem[{{Gaia Collaboration} {et~al.}(2016){Gaia Collaboration}, {Prusti},
  {de Bruijne}, {Brown}, {Vallenari}, {Babusiaux}, {Bailer-Jones}, {Bastian},
  {Biermann}, {Evans}, \& et~al.}]{Gaia2016}
{Gaia Collaboration}, {Prusti}, T., {de Bruijne}, J.~H.~J., {et~al.} 2016,
  \aap, 595, A1

\bibitem[{{Garc{\'{\i}}a} {et~al.}(2011){Garc{\'{\i}}a}, {Hekker}, {Stello},
  {Guti{\'e}rrez-Soto}, {Handberg}, {Huber}, {Karoff}, {Uytterhoeven},
  {Appourchaux}, {Chaplin}, {Elsworth}, {Mathur}, {Ballot},
  {Christensen-Dalsgaard}, {Gilliland}, {Houdek}, {Jenkins}, {Kjeldsen},
  {McCauliff}, {Metcalfe}, {Middour}, {Molenda-Zakowicz}, {Monteiro}, {Smith},
  \& {Thompson}}]{Garcia2011}
{Garc{\'{\i}}a}, R.~A., {Hekker}, S., {Stello}, D., {et~al.} 2011, \mnras, 414,
  L6

\bibitem[{{Gaulme} {et~al.}(2014){Gaulme}, {Jackiewicz}, {Appourchaux}, \&
  {Mosser}}]{Gaulme2014}
{Gaulme}, P., {Jackiewicz}, J., {Appourchaux}, T., \& {Mosser}, B. 2014, \apj,
  785, 5

\bibitem[{{Gaulme} {et~al.}(2016){Gaulme}, {McKeever}, {Jackiewicz}, {Rawls},
  {Corsaro}, {Mosser}, {Southworth}, {Mahadevan}, {Bender}, \&
  {Deshpande}}]{Gaulme2016}
{Gaulme}, P., {McKeever}, J., {Jackiewicz}, J., {et~al.} 2016, \apj, 832, 121

\bibitem[{{Gaulme} {et~al.}(2013){Gaulme}, {McKeever}, {Rawls}, {Jackiewicz},
  {Mosser}, \& {Guzik}}]{Gaulme2013}
{Gaulme}, P., {McKeever}, J., {Rawls}, M.~L., {et~al.} 2013, \apj, 767, 82

\bibitem[{{Goodman} \& {Weare}(2010)}]{Goodman2010}
{Goodman}, J. \& {Weare}, J. 2010, Communications in Applied Mathematics and
  Computational Science, Vol.~5, No.~1, p.~65-80, 2010, 5, 65

\bibitem[{{Guggenberger} {et~al.}(2017){Guggenberger}, {Hekker}, {Angelou},
  {Basu}, \& {Bellinger}}]{Guggenberger2017}
{Guggenberger}, E., {Hekker}, S., {Angelou}, G.~C., {Basu}, S., \& {Bellinger},
  E.~P. 2017, \mnras, 470, 2069

\bibitem[{{Guggenberger} {et~al.}(2016){Guggenberger}, {Hekker}, {Basu}, \&
  {Bellinger}}]{Guggenberger2016}
{Guggenberger}, E., {Hekker}, S., {Basu}, S., \& {Bellinger}, E. 2016, \mnras,
  460, 4277

\bibitem[{{Harvey}(1985)}]{Harvey1985}
{Harvey}, J. 1985, in ESA Special Publication, Vol. 235, Future Missions in
  Solar, Heliospheric \& Space Plasma Physics, ed. E.~{Rolfe} \& B.~{Battrick}

\bibitem[{{Hekker} {et~al.}(2011){Hekker}, {Elsworth}, {De Ridder}, {Mosser},
  {Garc{\'{\i}}a}, {Kallinger}, {Mathur}, {Huber}, {Buzasi}, {Preston}, {Hale},
  {Ballot}, {Chaplin}, {R{\'e}gulo}, {Bedding}, {Stello}, {Borucki}, {Koch},
  {Jenkins}, {Allen}, {Gilliland}, {Kjeldsen}, \&
  {Christensen-Dalsgaard}}]{Hekker2011a}
{Hekker}, S., {Elsworth}, Y., {De Ridder}, J., {et~al.} 2011, \aap, 525, A131

\bibitem[{{Herwig}(2000)}]{Herwig2000}
{Herwig}, F. 2000, \aap, 360, 952

\bibitem[{{Huber}(2015)}]{Huber2015}
{Huber}, D. 2015, in Astrophysics and Space Science Library, Vol. 408, Giants
  of Eclipse: The {$\zeta$} Aurigae Stars and Other Binary Systems, 169

\bibitem[{{Huber} {et~al.}(2011){Huber}, {Bedding}, {Stello}, {Hekker},
  {Mathur}, {Mosser}, {Verner}, {Bonanno}, {Buzasi}, {Campante}, {Elsworth},
  {Hale}, {Kallinger}, {Silva Aguirre}, {Chaplin}, {De Ridder},
  {Garc{\'{\i}}a}, {Appourchaux}, {Frandsen}, {Houdek}, {Molenda-{\.Z}akowicz},
  {Monteiro}, {Christensen-Dalsgaard}, {Gilliland}, {Kawaler}, {Kjeldsen},
  {Broomhall}, {Corsaro}, {Salabert}, {Sanderfer}, {Seader}, \&
  {Smith}}]{Huber2011}
{Huber}, D., {Bedding}, T.~R., {Stello}, D., {et~al.} 2011, \apj, 743, 143

\bibitem[{{Huber} {et~al.}(2017){Huber}, {Zinn}, {Bojsen-Hansen},
  {Pinsonneault}, {Sahlholdt}, {Serenelli}, {Silva Aguirre}, {Stassun},
  {Stello}, {Tayar}, {Bastien}, {Bedding}, {Buchhave}, {Chaplin}, {Davies},
  {Garc{\'{\i}}a}, {Latham}, {Mathur}, {Mosser}, \& {Sharma}}]{Huber2017}
{Huber}, D., {Zinn}, J., {Bojsen-Hansen}, M., {et~al.} 2017, \apj, 844, 102

\bibitem[{{Kallinger} {et~al.}(2018){Kallinger}, {Beck}, {Stello}, \&
  {Garcia}}]{Kallinger2018}
{Kallinger}, T., {Beck}, P.~G., {Stello}, D., \& {Garcia}, R.~A. 2018, \aap

\bibitem[{{Kallinger} {et~al.}(2014){Kallinger}, {De Ridder}, {Hekker},
  {Mathur}, {Mosser}, {Gruberbauer}, {Garc{\'{\i}}a}, {Karoff}, \&
  {Ballot}}]{Kallinger2014}
{Kallinger}, T., {De Ridder}, J., {Hekker}, S., {et~al.} 2014, \aap, 570, A41

\bibitem[{{Kallinger} {et~al.}(2010){Kallinger}, {Weiss}, {Barban}, {Baudin},
  {Cameron}, {Carrier}, {De Ridder}, {Goupil}, {Gruberbauer}, {Hatzes},
  {Hekker}, {Samadi}, \& {Deleuil}}]{Kallinger2010a}
{Kallinger}, T., {Weiss}, W.~W., {Barban}, C., {et~al.} 2010, \aap, 509, A77

\bibitem[{{Kjeldsen} \& {Bedding}(1995)}]{Kjeldsen1995}
{Kjeldsen}, H. \& {Bedding}, T.~R. 1995, \aap, 293, 87

\bibitem[{{Li} {et~al.}(2018){Li}, {Bedding}, {Li}, {Bi}, {Murphy}, {Corsaro},
  {Chen}, \& {Tian}}]{Li2018}
{Li}, Y., {Bedding}, T.~R., {Li}, T., {et~al.} 2018, \mnras, 476, 470

\bibitem[{{Lund} {et~al.}(2017){Lund}, {Silva Aguirre}, {Davies}, {Chaplin},
  {Christensen-Dalsgaard}, {Houdek}, {White}, {Bedding}, {Ball}, {Huber},
  {Antia}, {Lebreton}, {Latham}, {Handberg}, {Verma}, {Basu}, {Casagrande},
  {Justesen}, {Kjeldsen}, \& {Mosumgaard}}]{Lund2017}
{Lund}, M.~N., {Silva Aguirre}, V., {Davies}, G.~R., {et~al.} 2017, \apj, 835,
  172

\bibitem[{{Mathur} {et~al.}(2013){Mathur}, {Bruntt}, {Catala}, {Benomar},
  {Davies}, {Garc{\'{\i}}a}, {Salabert}, {Ballot}, {Mosser}, {R{\'e}gulo},
  {Chaplin}, {Elsworth}, {Handberg}, {Hekker}, {Mantegazza}, {Michel},
  {Poretti}, {Rainer}, {Roxburgh}, {Samadi}, {St{\c e}{\'s}licki},
  {Uytterhoeven}, {Verner}, {Auvergne}, {Baglin}, {Barcel{\'o} Forteza},
  {Baudin}, \& {Roca Cort{\'e}s}}]{Mathur2013}
{Mathur}, S., {Bruntt}, H., {Catala}, C., {et~al.} 2013, \aap, 549, A12

\bibitem[{{Miglio} {et~al.}(2012){Miglio}, {Brogaard}, {Stello}, {Chaplin},
  {D'Antona}, {Montalb{\'a}n}, {Basu}, {Bressan}, {Grundahl}, {Pinsonneault},
  {Serenelli}, {Elsworth}, {Hekker}, {Kallinger}, {Mosser}, {Ventura},
  {Bonanno}, {Noels}, {Silva Aguirre}, {Szabo}, {Li}, {McCauliff}, {Middour},
  \& {Kjeldsen}}]{Miglio2012}
{Miglio}, A., {Brogaard}, K., {Stello}, D., {et~al.} 2012, \mnras, 419, 2077

\bibitem[{{Miglio} {et~al.}(2013){Miglio}, {Chiappini}, {Morel}, {Barbieri},
  {Chaplin}, {Girardi}, {Montalb{\'a}n}, {Valentini}, {Mosser}, {Baudin},
  {Casagrande}, {Fossati}, {Silva Aguirre}, \& {Baglin}}]{Miglio2013}
{Miglio}, A., {Chiappini}, C., {Morel}, T., {et~al.} 2013, \mnras, 429, 423

\bibitem[{{Mosser} {et~al.}(2010){Mosser}, {Belkacem}, {Goupil}, {Miglio},
  {Morel}, {Barban}, {Baudin}, {Hekker}, {Samadi}, {De Ridder}, {Weiss},
  {Auvergne}, \& {Baglin}}]{Mosser2010}
{Mosser}, B., {Belkacem}, K., {Goupil}, M.-J., {et~al.} 2010, \aap, 517, A22

\bibitem[{{Mosser} {et~al.}(2012){Mosser}, {Goupil}, {Belkacem}, {Michel},
  {Stello}, {Marques}, {Elsworth}, {Barban}, {Beck}, {Bedding}, {De Ridder},
  {Garcia}, {Hekker}, {Kallinger}, {Samadi}, {Stumpe}, {Barclay}, \&
  {Burke}}]{Mosser2012b}
{Mosser}, B., {Goupil}, M.~J., {Belkacem}, K., {et~al.} 2012, VizieR Online
  Data Catalog, 354

\bibitem[{{Mosser} {et~al.}(2013){Mosser}, {Michel}, {Belkacem}, {Goupil},
  {Baglin}, {Barban}, {Provost}, {Samadi}, {Auvergne}, \&
  {Catala}}]{Mosser2013}
{Mosser}, B., {Michel}, E., {Belkacem}, K., {et~al.} 2013, \aap, 550, A126

\bibitem[{{Nieva} \& {Przybilla}(2012)}]{Nieva2012}
{Nieva}, M.-F. \& {Przybilla}, N. 2012, \aap, 539, A143

\bibitem[{{Pande} {et~al.}(2018){Pande}, {Bedding}, {Huber}, \&
  {Kjeldsen}}]{Pande2018}
{Pande}, D., {Bedding}, T.~R., {Huber}, D., \& {Kjeldsen}, H. 2018, \mnras,
  480, 467

\bibitem[{{Paxton} {et~al.}(2011){Paxton}, {Bildsten}, {Dotter}, {Herwig},
  {Lesaffre}, \& {Timmes}}]{Paxton2011}
{Paxton}, B., {Bildsten}, L., {Dotter}, A., {et~al.} 2011, \apjs, 192, 3

\bibitem[{{Paxton} {et~al.}(2018){Paxton}, {Schwab}, {Bauer}, {Bildsten},
  {Blinnikov}, {Duffell}, {Farmer}, {Goldberg}, {Marchant}, {Sorokina},
  {Thoul}, {Townsend}, \& {Timmes}}]{Paxton2018}
{Paxton}, B., {Schwab}, J., {Bauer}, E.~B., {et~al.} 2018, \apjs, 234, 34

\bibitem[{{P{\'e}rez Hern{\'a}ndez} {et~al.}(2016){P{\'e}rez Hern{\'a}ndez},
  {Garc{\'\i}a}, {Corsaro}, {Triana}, \& {De Ridder}}]{PerezHernandez2016}
{P{\'e}rez Hern{\'a}ndez}, F., {Garc{\'\i}a}, R.~A., {Corsaro}, E., {Triana},
  S.~A., \& {De Ridder}, J. 2016, \aap, 591, A99

\bibitem[{{Przybilla} {et~al.}(2013){Przybilla}, {Nieva}, {Irrgang}, \&
  {Butler}}]{Przybilla2013}
{Przybilla}, N., {Nieva}, M.~F., {Irrgang}, A., \& {Butler}, K. 2013, in EAS
  Publications Series, Vol.~63, EAS Publications Series, ed. G.~{Alecian},
  Y.~{Lebreton}, O.~{Richard}, \& G.~{Vauclair}, 13--23

\bibitem[{{Rawls}(2016)}]{Rawls2016b}
{Rawls}, M.~L. 2016, PhD thesis, New Mexico State University

\bibitem[{{Rawls} {et~al.}(2016){Rawls}, {Gaulme}, {McKeever}, {Jackiewicz},
  {Orosz}, {Corsaro}, {Beck}, {Mosser}, {Latham}, \& {Latham}}]{Rawls2016a}
{Rawls}, M.~L., {Gaulme}, P., {McKeever}, J., {et~al.} 2016, \apj, 818, 108

\bibitem[{{Ricker} {et~al.}(2015){Ricker}, {Winn}, {Vanderspek}, {Latham},
  {Bakos}, {Bean}, {Berta-Thompson}, {Brown}, {Buchhave}, {Butler}, {Butler},
  {Chaplin}, {Charbonneau}, {Christensen-Dalsgaard}, {Clampin}, {Deming},
  {Doty}, {De Lee}, {Dressing}, {Dunham}, {Endl}, {Fressin}, {Ge}, {Henning},
  {Holman}, {Howard}, {Ida}, {Jenkins}, {Jernigan}, {Johnson}, {Kaltenegger},
  {Kawai}, {Kjeldsen}, {Laughlin}, {Levine}, {Lin}, {Lissauer}, {MacQueen},
  {Marcy}, {McCullough}, {Morton}, {Narita}, {Paegert}, {Palle}, {Pepe},
  {Pepper}, {Quirrenbach}, {Rinehart}, {Sasselov}, {Sato}, {Seager},
  {Sozzetti}, {Stassun}, {Sullivan}, {Szentgyorgyi}, {Torres}, {Udry}, \&
  {Villasenor}}]{Ricker2015}
{Ricker}, G.~R., {Winn}, J.~N., {Vanderspek}, R., {et~al.} 2015, Journal of
  Astronomical Telescopes, Instruments, and Systems, 1, 014003

\bibitem[{{Rodrigues} {et~al.}(2017){Rodrigues}, {Bossini}, {Miglio},
  {Girardi}, {Montalb{\'a}n}, {Noels}, {Trabucchi}, {Coelho}, \&
  {Marigo}}]{Rodrigues2017}
{Rodrigues}, T.~S., {Bossini}, D., {Miglio}, A., {et~al.} 2017, \mnras, 467,
  1433

\bibitem[{{Salgado} {et~al.}(2017){Salgado}, {Gonz{\'a}lez-N{\'u}{\~n}ez},
  {Guti{\'e}rrez-S{\'a}nchez}, {Segovia}, {Dur{\'a}n}, {Hern{\'a}ndez}, \&
  {Arviset}}]{Salgado2017}
{Salgado}, J., {Gonz{\'a}lez-N{\'u}{\~n}ez}, J., {Guti{\'e}rrez-S{\'a}nchez},
  R., {et~al.} 2017, Astronomy and Computing, 21, 22

\bibitem[{{Samadi} {et~al.}(2015){Samadi}, {Belkacem}, \& {Sonoi}}]{Samadi2015}
{Samadi}, R., {Belkacem}, K., \& {Sonoi}, T. 2015, in EAS Publications Series,
  Vol. 73-74, EAS Publications Series, 111--191

\bibitem[{{Sharma} {et~al.}(2016){Sharma}, {Stello}, {Bland-Hawthorn}, {Huber},
  \& {Bedding}}]{Sharma2016}
{Sharma}, S., {Stello}, D., {Bland-Hawthorn}, J., {Huber}, D., \& {Bedding},
  T.~R. 2016, \apj, 822, 15

\bibitem[{{Tassoul}(1980)}]{Tassoul1980}
{Tassoul}, M. 1980, \apjs, 43, 469

\bibitem[{{Theme{\ss}l} {et~al.}(2018){Theme{\ss}l}, {Hekker}, {Southworth},
  {Beck}, {Pavlovski}, {Tkachenko}, {Angelou}, {Ball}, {Barban}, {Corsaro},
  {Elsworth}, {Handberg}, \& {Kallinger}}]{Themessl2018}
{Theme{\ss}l}, N., {Hekker}, S., {Southworth}, J., {et~al.} 2018, \mnras, 478,
  4669

\bibitem[{{Torres} {et~al.}(2010){Torres}, {Andersen}, \&
  {Gim{\'e}nez}}]{Torres2010}
{Torres}, G., {Andersen}, J., \& {Gim{\'e}nez}, A. 2010, \aapr, 18, 67

\bibitem[{{Townsend} \& {Teitler}(2013)}]{Townsend2013}
{Townsend}, R.~H.~D. \& {Teitler}, S.~A. 2013, \mnras, 435, 3406

\bibitem[{{Vrard} {et~al.}(2018){Vrard}, {Kallinger}, {Mosser}, {Barban},
  {Baudin}, {Belkacem}, \& {Cunha}}]{Vrard2018}
{Vrard}, M., {Kallinger}, T., {Mosser}, B., {et~al.} 2018, \aap, 616, A94

\bibitem[{{White} {et~al.}(2011){White}, {Bedding}, {Stello},
  {Christensen-Dalsgaard}, {Huber}, \& {Kjeldsen}}]{White2011}
{White}, T.~R., {Bedding}, T.~R., {Stello}, D., {et~al.} 2011, \apj, 743, 161

\bibitem[{{White} {et~al.}(2017){White}, {Benomar}, {Silva Aguirre}, {Ball},
  {Bedding}, {Chaplin}, {Christensen-Dalsgaard}, {Garcia}, {Gizon}, {Stello},
  {Aigrain}, {Antia}, {Appourchaux}, {Bazot}, {Campante}, {Creevey}, {Davies},
  {Elsworth}, {Gaulme}, {Handberg}, {Hekker}, {Houdek}, {Howe}, {Huber},
  {Karoff}, {Marques}, {Mathur}, {McQuillan}, {Metcalfe}, {Mosser}, {Nielsen},
  {R{\'e}gulo}, {Salabert}, \& {Stahn}}]{White2017}
{White}, T.~R., {Benomar}, O., {Silva Aguirre}, V., {et~al.} 2017, \aap, 601,
  A82

\end{thebibliography}

\clearpage
\newpage
\begin{appendix}
\section{MESA inlist file}
\label{sec: MESAInlist}

The running and output of the \textsc{mesa} stellar evolutionary code is configured by an inlist file. While the code does have a default configuration, any parameters that the user wishes to alter must be specified in the inlist file. The contents of the inlist file for creating our RG models are shown below:

\small\begin{verbatim}
&star_job
    
    show_log_description_at_start = .false.
    
    load_saved_model = .false.
    create_pre_main_sequence_model = .true.
    
    kappa_file_prefix = 'OP_a09_p13'
    kappa_lowT_prefix = 'lowT_fa05_a09p'
    kappa_CO_prefix = 'a09_p13_co'
    
    initial_zfracs = 8
    
    change_net = .true.
    new_net_name = 
    'pp_cno_extras_o18_ne22_extraiso.net'
    change_initial_net = .true.
    
    pgstar_flag = .false.
    
    pause_before_terminate = .false.
    
    save_photo_when_terminate = .false.
    save_model_when_terminate = .false.
    write_profile_when_terminate = .false.
    save_pulse_data_when_terminate = .false.
    
    new_rotation_flag = .false.
    change_rotation_flag = .false.
    change_initial_rotation_flag = .false.
    
    new_omega = 0
    set_omega = .false.
    set_initial_omega = .false.
    
/ !end of star_job namelist

&controls

    star_history_name = 
    initial_mass = 
    mixing_length_alpha = 1.8
    overshoot_f_above_burn_h_core = 0.02
    overshoot_f_above_burn_he_core = 0.02
    min_D_mix = 1.0
    
    initial_y = 0.2485
    initial_z = 0.018
    
    varcontrol_target = 1d-4
    
    log_directory = 
    
    terminal_interval = 200
    
    photo_interval = 1000000
    photo_directory = './photos'
    
    write_profiles_flag = .false.
    
    history_interval = 1
    
    write_pulse_data_with_profile = .true.
    pulse_data_format = 'GYRE'
    add_atmosphere_to_pulse_data = .true.
    add_center_point_to_pulse_data = .true.
    keep_surface_point_for_pulse_data = .true.
    add_double_points_to_pulse_data = .true.
    interpolate_rho_for_pulse_data = .true.
    threshold_grad_mu_for_double_point = 5d0
    
    alpha_bdy_core_overshooting = 5
    he_core_boundary_h1_fraction = 1d-2
    
    hot_wind_scheme = 'Vink'
    Vink_scaling_factor = 0.3d0
    hot_wind_full_on_T  = 1d4
    
    xa_central_lower_limit_species(1) = 'he4'
    xa_central_lower_limit(1) = 1d-3
    
    remove_small_D_limit = 1d-6
    use_Ledoux_criterion = .true.
    
    num_cells_for_smooth_gradL_composition_term = 0
    
    alpha_semiconvection = 0d0
    semiconvection_option = 
    'Langer_85 mixing; gradT = gradr'
    thermohaline_coeff = 0d0
    alt_scale_height_flag = .true.
    MLT_option = 'Cox'
    mlt_gradT_fraction = -1
    okay_to_reduce_gradT_excess = .false.
    
    set_min_D_mix = .true.
    
    D_mix_ov_limit = 5d-2
    max_brunt_B_for_overshoot = 0
    limit_overshoot_Hp_using_size_of_convection_zone = 
    .true.
    overshoot_alpha = -1
    
    predictive_mix(1) = .true.
    predictive_zone_type(1) = 'burn_H'
    predictive_zone_loc(1) = 'core'
    predictive_bdy_loc(1) = 'any'
    
    predictive_mix(2) = .true.
    predictive_zone_type(2) = 'burn_He'
    predictive_zone_loc(2) = 'core'
    predictive_bdy_loc(2) = 'any'
    
    predictive_mix(3) = .true.
    predictive_zone_type(3) = 'nonburn'
    predictive_zone_loc(3) = 'shell'
    predictive_bdy_loc(3) = 'any'
    
    predictive_mix(4) = .true.
    predictive_zone_type(4) = 'burn_H'
    predictive_zone_loc(4) = 'shell'
    predictive_bdy_loc(4) = 'any'
    
    conv_bdy_mix_softening_f0 = 0.002
    conv_bdy_mix_softening_f = 0.001
    conv_bdy_mix_softening_min_D_mix = 1d-1
    
    overshoot_f0_above_burn_h_core = 0.001
    
    overshoot_f0_above_burn_h_shell = 0.001
    overshoot_f_above_burn_h_shell  = 0.005
    overshoot_f0_below_burn_h_shell = 0.001
    overshoot_f_below_burn_h_shell  = 0.005
    
    overshoot_f0_above_burn_he_core = 0.001
    
    overshoot_f0_above_nonburn_shell = 0.001
    overshoot_f_above_nonburn_shell  = 0.001
    overshoot_f0_below_nonburn_shell = 0.005
    overshoot_f_below_nonburn_shell  = 0.005
    
    smooth_convective_bdy = .false.
    
    do_element_diffusion = .false.
    
    which_atm_option = 'simple_photosphere'
    
    cubic_interpolation_in_X = .false.
    cubic_interpolation_in_Z = .false.
    
    num_cells_for_smooth_brunt_B = 0
    interpolate_rho_for_pulsation_info = .true.
    
    max_allowed_nz = 40000
      
    mesh_delta_coeff = 0.4
    mesh_adjust_use_quadratic = .true.
    mesh_adjust_get_T_from_E = .true.
    
    P_function_weight = 40
    T_function1_weight = 110 
    
    T_function2_weight = 0
    T_function2_param = 2d4
    
    gradT_function_weight = 0
    
    xtra_coef_os_above_burn_h = 0.1d0
    xtra_dist_os_above_burn_h = 2d0
    
    mesh_dlogX_dlogP_extra = 0.15                   
    mesh_dlogX_dlogP_full_on = 1d-6                 
    mesh_dlogX_dlogP_full_off = 1d-12               
    mesh_logX_species(1) = 'he4'                      
      
    
    xtra_coef_czb_full_on = 1.0d0                     
    xtra_coef_czb_full_off = 1.0d0                    
    
    xtra_coef_a_l_hb_czb = 0.5d0                      
    xtra_dist_a_l_hb_czb = 1d0                        
    xtra_coef_b_l_hb_czb = 0.5d0                      
    xtra_dist_b_l_hb_czb = 1d0                      
    
    xtra_coef_a_l_hb_czb = 0.5d0                    
    xtra_dist_a_l_hb_czb = 1d0                      
    xtra_coef_b_l_hb_czb = 0.5d0                    
    xtra_dist_b_l_hb_czb = 1d0                      
    
    ! non-burning zone
    xtra_coef_a_l_nb_czb = 0.5d0                   
    xtra_dist_a_l_nb_czb = 1d0                    
    xtra_coef_b_l_nb_czb = 0.5d0                   
    xtra_dist_b_l_nb_czb = 1d0                     
    
    xtra_coef_a_l_nb_czb = 0.5d0                 
    xtra_dist_a_l_nb_czb = 1d0                    
    xtra_coef_b_l_nb_czb = 0.5d0                  
    xtra_dist_b_l_nb_czb = 1d0                   
    
    ! He burning zone
    xtra_coef_a_l_heb_czb = 0.5d0               
    xtra_dist_a_l_heb_czb = 1d0                 
    xtra_coef_b_l_heb_czb = 0.5d0               
    xtra_dist_b_l_heb_czb = 1d0                
    
    xtra_coef_a_l_heb_czb = 0.5d0          
    xtra_dist_a_l_heb_czb = 1d0            
    xtra_coef_b_l_heb_czb = 0.5d0          
    xtra_dist_b_l_heb_czb = 1d0         
    
    xtra_coef_os_full_on = 1.0d0
    xtra_coef_os_full_off = 1.0d0
    
    xtra_coef_os_above_burn_h = 0.5d0
    xtra_dist_os_above_burn_h = 0.5d0
    xtra_coef_os_below_burn_h = 0.5d0
    xtra_dist_os_below_burn_h = 0.5d0
    
    xtra_coef_os_above_nonburn = 0.5d0
    xtra_dist_os_above_nonburn = 0.5d0
    xtra_coef_os_below_nonburn = 0.5d0
    xtra_dist_os_below_nonburn = 0.5d0
    
    xtra_coef_os_above_burn_he = 0.5d0
    xtra_dist_os_above_burn_he = 0.5d0
    xtra_coef_os_below_burn_he = 0.5d0
    xtra_dist_os_below_burn_he = 0.5d0
    
/ ! end of controls namelist
\end{verbatim}
\normalsize

\section{GYRE inlist file}
\label{sec: GYREInlist}

Similar to \textsc{mesa} stellar evolutionary code, the running and output of the \textsc{gyre} stellar pulsation code is configured by an inlist file. The code will output the pulsational parameters of the modes specified in the input file for a specific input model. The contents of the inlist file for computing the pulsational frequencies and parameters for our input RG models are shown below:

\small\begin{verbatim}
&constants
/

&model
    
    model_type = 'EVOL'
    file = 
    file_format = 'MESA'
    
/

&mode
    l = 0
    m = 0
    tag = 'l0m0' ! Tag for namelist matching

/

&mode
    l = 1
    m = 0
    tag = 'l1m0' ! Tag for namelist matching

/

&mode
    l = 2
    m = 0
    tag = 'l2m0' ! Tag for namelist matching

/

&mode
    l = 3
    m = 0
    tag = 'l3m0'

/

&osc
    inner_bound = 'REGULAR'
    outer_bound = 'JCD'
    variables_set = 'JCD'
    inertia_norm = 'BOTH'
    rotation_method = 'NULL'
/

&num
    diff_scheme = 'MAGNUS_GL4'
    n_iter_max = 50
/

&scan
    grid_type = 'LINEAR'
    grid_frame = 'INERTIAL'
    freq_min = 
    freq_max = 
    freq_min_units = 'UHZ'
    freq_max_units = 'UHZ'
    freq_frame = 'INERTIAL'
    n_freq = 400
    
/

&grid
    alpha_osc = 10		
    alpha_exp = 5
    n_inner = 5
    alpha_thm = 0
    alpha_str = 0

/

&ad_output
    summary_file = 'Mini0100_profiles/
    Mini0100_at_nu_max_0020.profile.freqs'
    freq_units = 'UHZ'
    summary_file_format = 'TXT'
    summary_item_list = 'l,m,n_p,n_g,n_pg,freq,E_norm'
/

&nad_output
/

\end{verbatim}
\normalsize

\newpage
\section{The variation of the normalised mode inertia with frequency}
\label{sec: ModeInertia}

\vspace{-10pt}

\begin{figure}[H]
\centering
\includegraphics[width=\hsize]{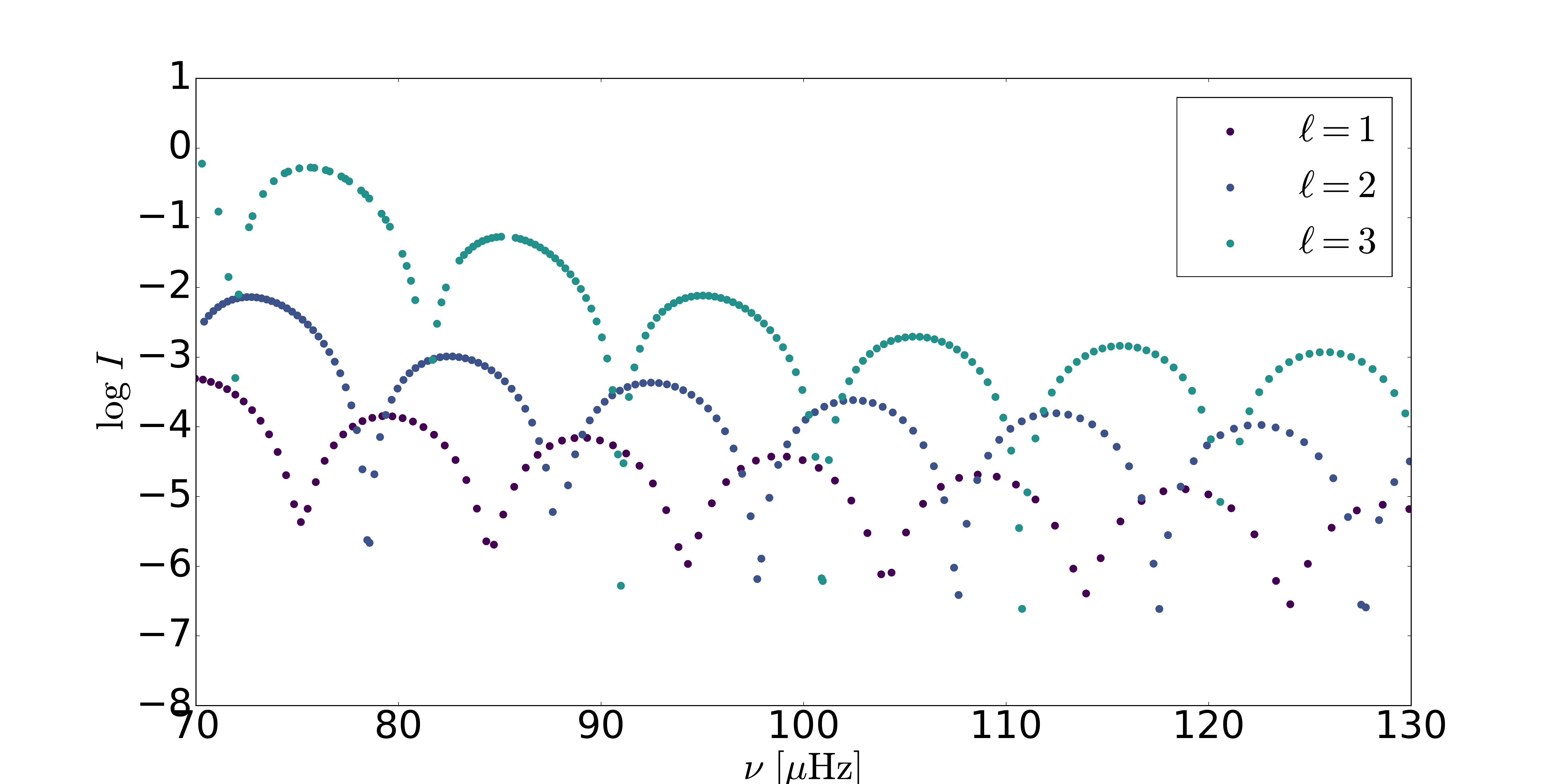}
\caption[The variation of the normalised mode inertia with frequency.]{The variation of the normalised mode inertia values for each frequency output by \textsc{gyre} for a star with $\nu_{\text{max}}\sim100$ $\mu$Hz. The $\ell=1$, 2 and 3 modes are colour-coded as displayed the legend. It should be noted that the local minima in mode inertia in intervals of $\Delta\nu$, and the general decreasing trend of mode inertia with frequency.}
\label{fig: MI}
\end{figure}

\section{Power spectrum fit examples}
\label{sec: powerspec}

This section contains examples of the fits obtained for our synthetic single- and binary-RG PSDs.

\begin{figure}[H]
\centering
\includegraphics[width=\hsize]{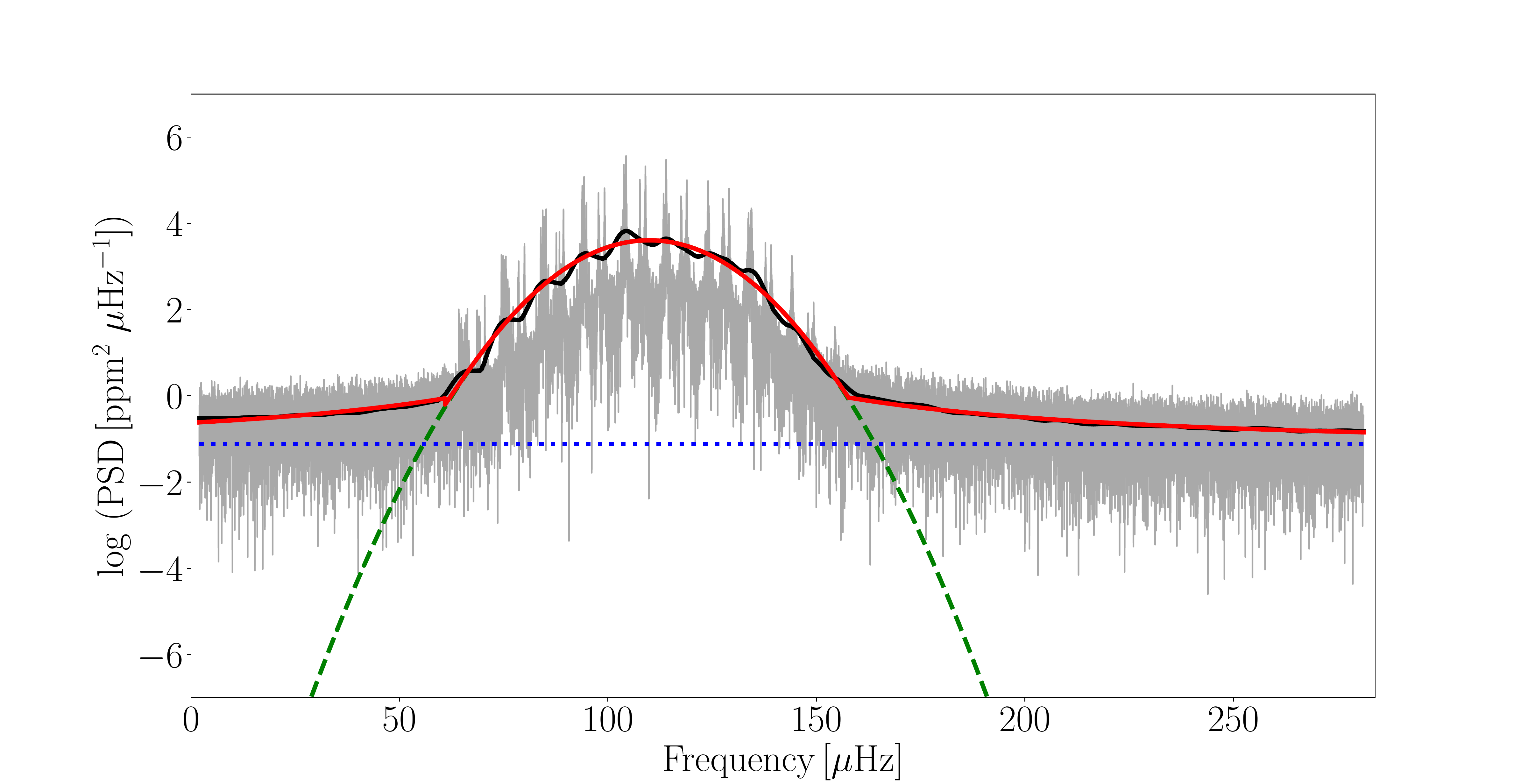}
\caption[An example of a single-RG PSD fit where the light curve contains only the pulsational signal of the RG.]{An example of a single-RG PSD fit where light curve contains only the pulsational signal of the RG. The grey line is the unsmoothed PSD, the black line is the smoothed PSD, the red line represents the overall fit, the dashed green line represents the Gaussian used to fit the pulsational power excess, and the blue dotted line represents the white noise.}
\label{fig: singlepuls_fitex}

\centering
\includegraphics[width=\hsize]{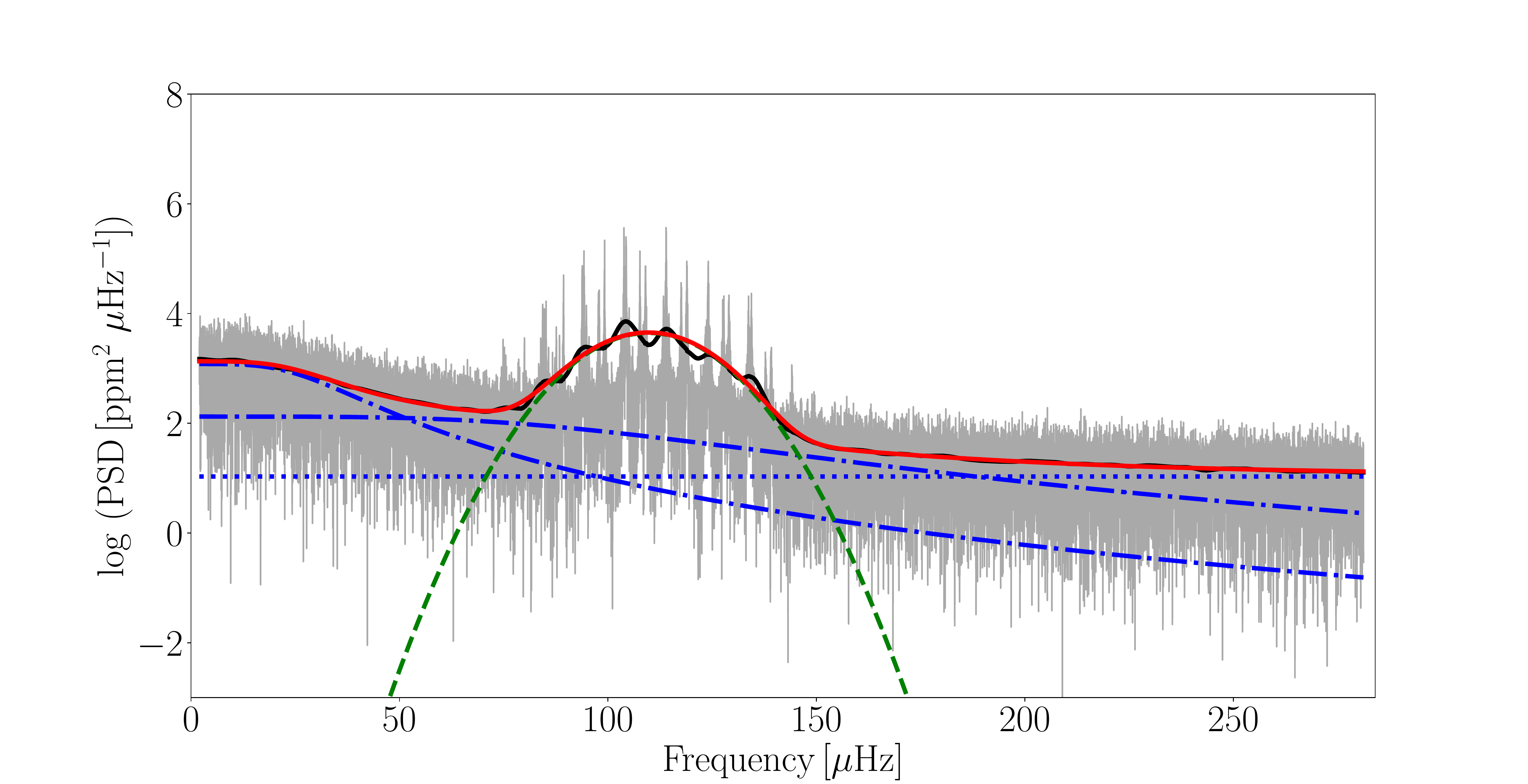}
\caption[An example of a single-RG PSD fit where the light curve contains both the pulsational and granulation signals of the RG.]{An example of a single-RG PSD fit where the light curve contains both the pulsational and granulation signals of the RG. The grey, black, red, green and blue lines represent the same signals as in Figure \ref{fig: singlepuls_fitex}, with the addition of blue dash-dotted lines to represent the super-Lorentzians.}
\label{fig: singlegran_fitex}
\end{figure}
\begin{figure}[H]
\centering
\includegraphics[width=\hsize]{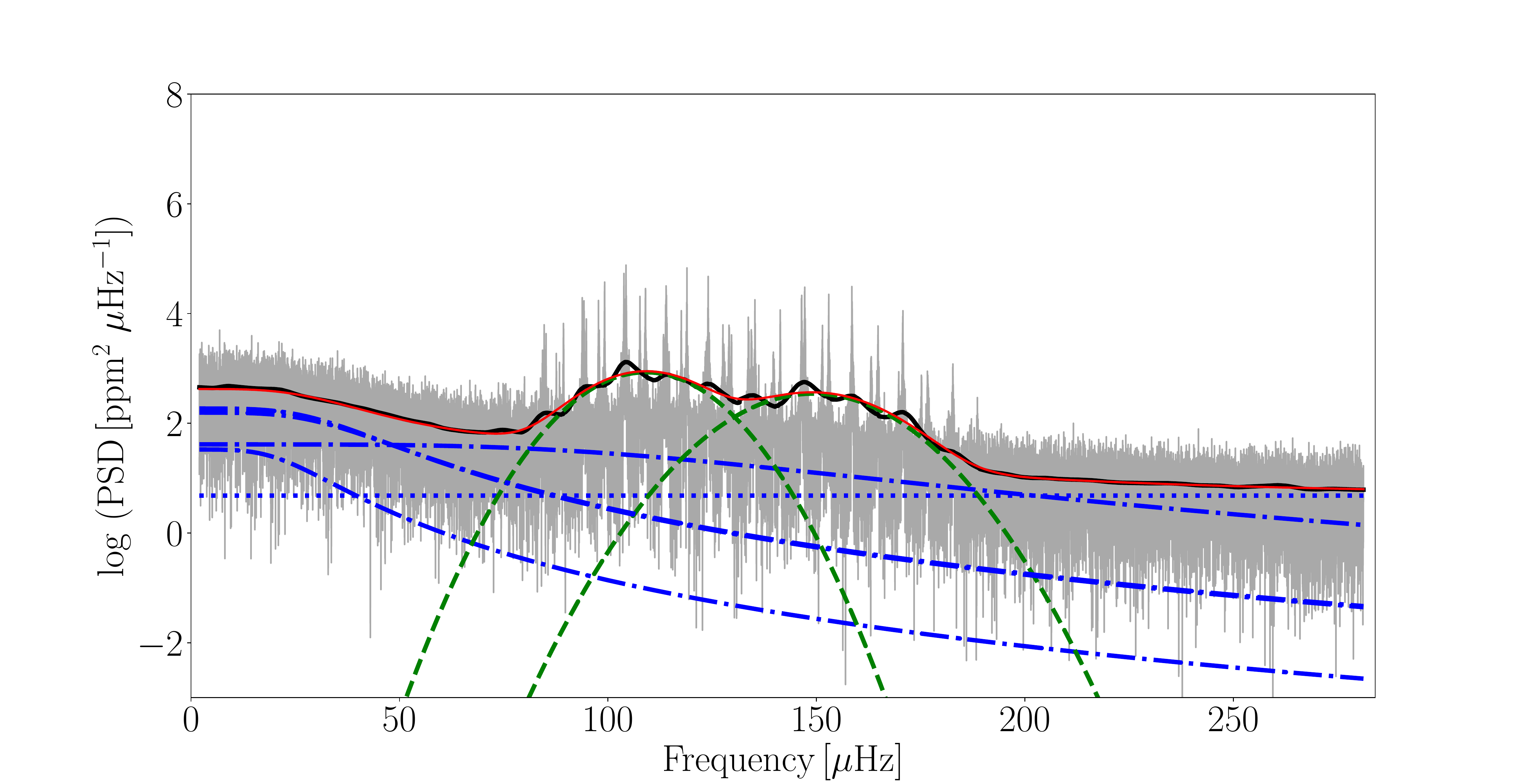}
\caption[An example of a RG-binary PSD fit where both components, with very similar $\nu_{\text{max}}$ values, are pulsating.]{An example of a RG-binary PSD fit where both components, with very similar $\nu_{\text{max}}$ values, are pulsating. Here we use four super-Lorentzians to fit the granulation signal in the PSD. The grey, black, red, green and blue lines (both dashed and dotted) represent the same signals as in Figure \ref{fig: singlegran_fitex}.}
\label{fig: binoverlap_fitex}

\centering
\includegraphics[width=\hsize]{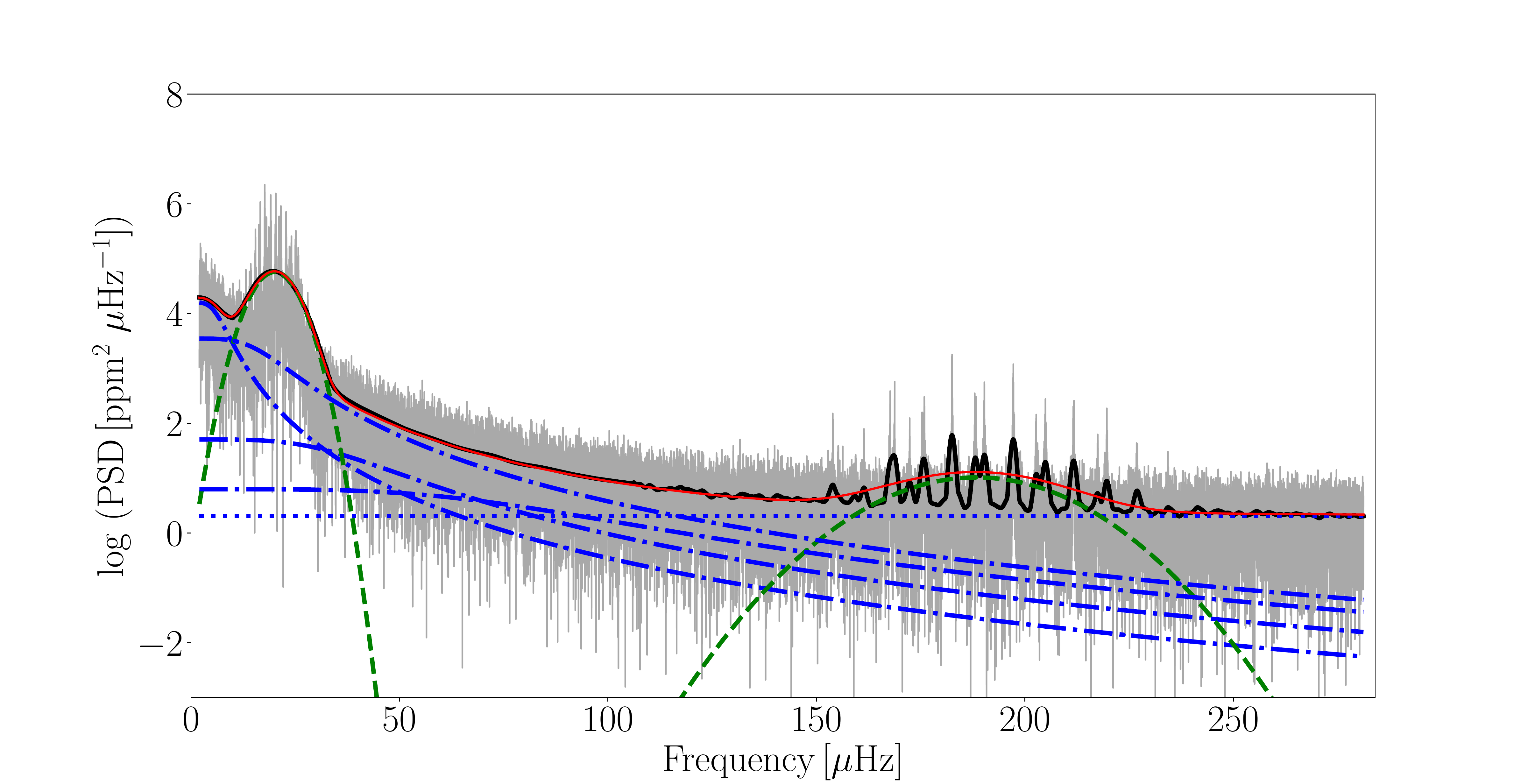}
\caption[An example of a RG-binary PSD fit where both components, with very different $\nu_{\text{max}}$ values, are pulsating.]{An example of a RG-binary PSD fit where both components, with very different $\nu_{\text{max}}$ values, are pulsating. Here we use four super-Lorentzians to fit the granulation signal in the PSD. The grey, black, red, green and blue lines (both dashed and dotted) represent the same signals as in Figure \ref{fig: singlegran_fitex}.}
\label{fig: binseparated_fitex}

\centering
\includegraphics[width=\hsize]{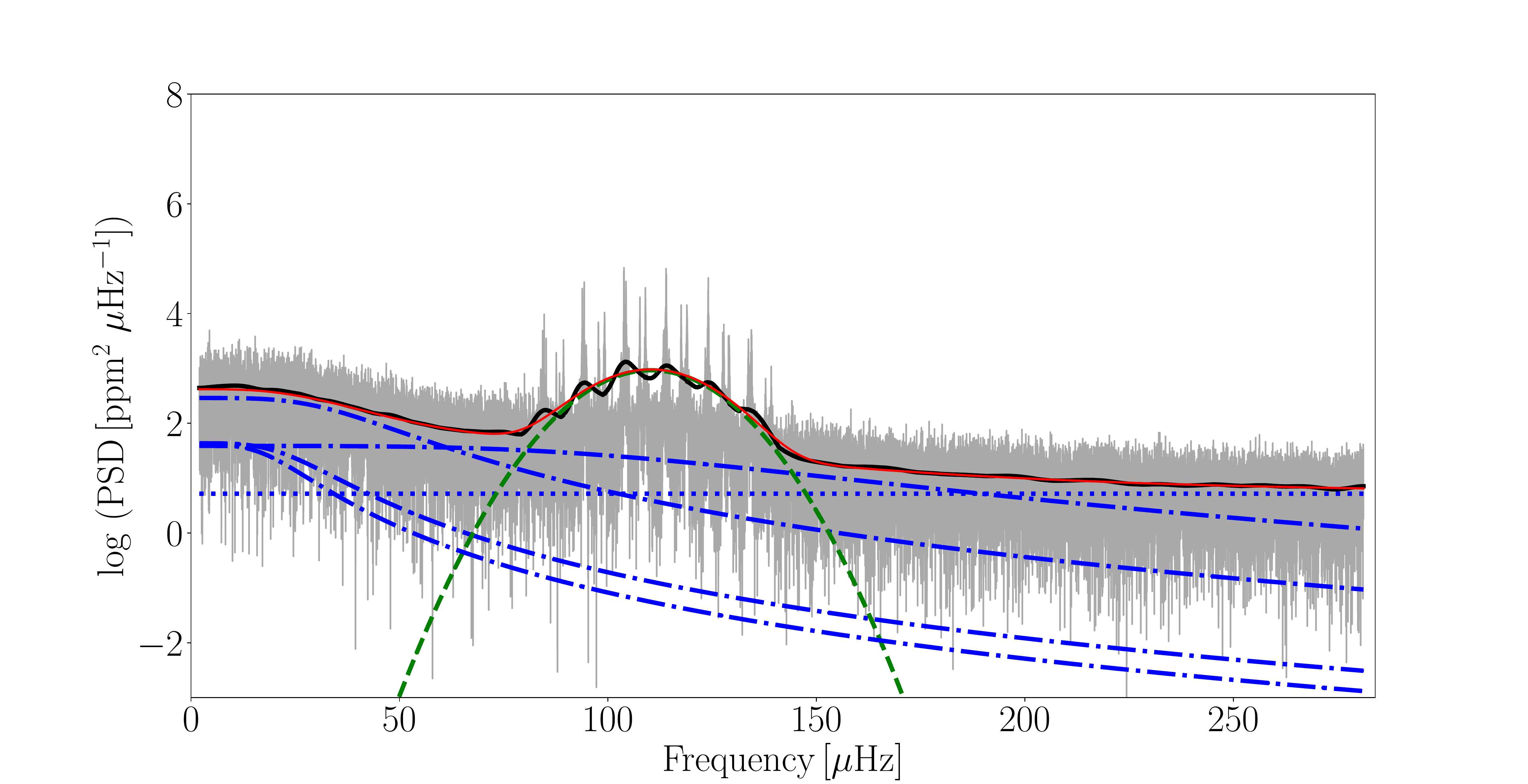}
\caption[An example of a RG-binary PSD fit where only one component is pulsating.]{An example of a RG-binary PSD fit where only one component is pulsating. Here we use four super-Lorentzians to fit the granulation signal in the PSD. The grey, black, red, green and blue lines (both dashed and dotted) represent the same signals as in Figure \ref{fig: singlegran_fitex}.}
\label{fig: bin1puls_fitex}
\end{figure}
\end{appendix}

\end{document}